# High-Performance Ferroelectric Field-Effect Transistors with Ultra-High Current and Carrier Densities

Seunguk Song,[1,6] Kwan-Ho Kim,[1,6] Rachael Keneipp,[2] Nicholas Trainor,[3] Chen Chen,[4] Jeffrey Zheng,[5] Joan M. Redwing,[3,4] Marija Drndić,[2] Roy H. Olsson III,[1] and Deep Jariwala[1,*]

[1]*Department of Electrical and Systems Engineering, University of Pennsylvania, Philadelphia, Pennsylvania 19104, United States*

[2]*Department of Physics and Astronomy, University of Pennsylvania, Philadelphia, Pennsylvania 19104, United States*

[3]*Department of Materials and Science and Engineering, Pennsylvania State University, University Park, Pennsylvania 16801, United States*

[4]*2D Crystal Consortium Materials Innovation Platform, Materials Research Institute, Pennsylvania State University, University Park, Pennsylvania, 16801 United States*

[5]*Department of Materials Science and Engineering, University of Pennsylvania, Philadelphia, Pennsylvania 19104, United States.*

[6]*These authors equally contributed to this work.*

[*]Author to whom correspondence should be addressed: dmj@seas.upenn.edu

## ABSTRACT

Ferroelectric field-effect transistors (FeFET) with two-dimensional (2D) semiconductor channels are promising low-power, embedded non-volatile memory (NVM) candidates for next-generation in-memory computing. However, the performance of FeFETs can be limited by a charge imbalance between the ferroelectric layer and the channel, and for low-dimensional semiconductors, also by a high contact resistance between the metal electrodes and the channel. Here, we report a significant enhancement in performance of contact-engineered FeFETs with a 2D $MoS_2$ channel and a ferroelectric $Al_{0.68}Sc_{0.32}N$ (AlScN) gate dielectric. Replacing Ti with In contact electrodes results in a fivefold increase in on-state current (~120 μA/μm at 1 V) and on-to-off ratio (~$2\cdot10^7$) in the FeFETs. In addition, the high carrier concentration in the $MoS_2$ channel during the on-state (> $10^{14}$ cm$^{-2}$) facilitates the observation of a metal-to-insulator phase transition in monolayer $MoS_2$ permitting observation of high field effect mobility (> 100 cm$^2$V$^{-1}$s$^{-1}$) at cryogenic temperatures. Our work and devices broaden the potential of FeFETs and provides a unique platform to implement high-carrier-density transport in a 2D channel.

## INTRODUCTION

The ferroelectric field effect transistor (FeFET) device has garnered attention due to its desirable properties such as a compact, non-volatile memory, high on-to-off current ratio, high access speed, and low programming energy.[1] However, the majority of FeFETs employing conventional semiconductor channels, such as Si, face a significant challenge due to the charge imbalance between the ferroelectric layer and the channel.[2,3] For example, when $Hf_xZr_{1-x}O_2$, which shows a remnant polarization ($P_r$) of ~10–30 μC/cm$^2$, is used in the gate stack, the carrier density in the Si channel should ideally reach ~$10^{14}$ cm$^{-2}$ to achieve a charge balance with the $P_r$ of $Hf_xZr_{1-x}O_2$. However, the carrier concentration in conventional Si channel FeFETs can only reach up to 1–$2\times10^{13}$ cm$^{-2}$ due to charge screening by the native $SiO_x$ interface and the 3D nature of the semiconductor.[2,3] This results in a large discrepancy in charge density between the channel and $Hf_xZr_{1-x}O_2$.[2] This mismatch causes partial polarization of the FeFETs and the accumulation of trapped carriers at the Si/$Hf_xZr_{1-x}O_2$ interface, resulting in various reliability issues.[2,4,5]. On the other hand, several reports have highlighted the doping capacity of 2D materials, particularly $MoS_2$, in supporting carrier concentration exceeding $10^{14}$–$10^{15}$ cm$^{-2}$.[6-9] This is particularly advantageous to mitigate the charge imbalance issue between the ferroelectric and 2D channel material. The substantial capability of 2D $MoS_2$ to support higher carrier concentrations owing to its 2D density of states further makes them promising channel material candidates, especially when paired with ferroelectric materials with large $P_r$ in the gate stack.[8,10-15]

Recently, $Al_{0.68}Sc_{0.32}N$ (AlScN) has been explored as the ferroelectric material in 2D FeFETs due to its high $P_r$ value (~80-115 μC/cm$^2$) and the feasibility of using a low thermal budget process for back-end-of-line direct integration.[16,17] The high $P_r$ of AlScN, coupled with the high doping capacity and ultraclean van der Waals interface of 2D $MoS_2$, can offer an excellent combination for high-performance FeFETs.[17,18] Specifically, the carrier density induced by the $P_r$ of AlScN is theoretically ~$5$–$7\times10^{14}$ cm$^{-2}$ ($=P_r/q$, where $q = 1.6 \times 10^{-19}$ C is the elementary charge),[19] matching well with the doping capacity of $MoS_2$ (~$10^{14}$–$10^{15}$ cm$^{-2}$).[6-9] However, the reported $MoS_2$/AlScN FeFETs[17,18] may not have fully taken advantage of the high $P_r$ of AlScN due to imperfect metal-semiconductor contacts.[20-24] For example, high contact resistance results in degradation in sheet conductance, transconductance, and on-state ($I_{on}$) and off-state currents ($I_{off}$), ultimately impeding the performance of ultra-scaled 2D devices.[20-24] Therefore, to fully utilize the high $P_r$ of AlScN, it is important to tackle the persistent challenge of high contact resistance between the 2D channel and contact electrodes. This entails resolving issues such as energy level mismatch at the metal-semiconductor junction (MSJ) interface,[20,21] high-energy metal deposition processes[22], and chemical reactions of metals with 2D semiconducting transition metal dichalcogenides (TMDs).[20,24] Moreover, in the case of 2D FeFETs, there have been no systematic investigations into the effects of contact interface on key performance indicators such as $I_{on}/I_{off}$, and programming/erasing voltage.[8] This is particularly significant because low conductance may result in low read current and slow write/erase speed at the same read voltage.[2,4,5,8] Furthermore, ensuring scalable production of low-resistance contact electrodes is crucial for the high reproducibility and reliability of 2D-channel/ferroelectric FeFETs for eventual production and commercialization.

Here, we report the high current and high carrier density in In-contact based $MoS_2$ channel FeFETs with AlScN as the ferroelectric. We observe the $I_{on}$ of In-contacted $MoS_2$/AlScN FeFETs approaching ~300 (~120) μA/μm at a drain-to-source voltage ($V_{ds}$) of 3 (1) V, resulting in a high $I_{on}/I_{off}$ of $\sim 2 \cdot 10^7$ on average, which is ~3 times higher than with Ti contacts. Remarkably, $I_{on}$ (~100 μA/μm), $I_{on}/I_{off}$ ($\sim 2 \cdot 10^7$), and memory windows normalized by the ferroelectric thickness (~0.11 V/nm) of the $MoS_2$/AlScN FeFETs surpass those reported for any other 2D-channel/ferroelectric FeFETs. In conjunction with the negligible thermal emission barrier height of the In-$MoS_2$ MSJ interface, the high $P_r$ of the AlScN enables a significantly large number of carriers to be injected into the monolayer channel ($> 4.8 \times 10^{14}$ $cm^{-2}$) with a high field-effect mobility (~31 $cm^2V^{-1}s^{-1}$). In addition, the large doping capacity enables the observation of a metal-to-insulator phase transition in monolayer $MoS_2$, concurrently with a band-like transport as observed in variable temperature measurements further suggesting high doping levels enabled by the high $P$r of AlScN. Finally, we have also explored the pulse dynamic characteristics and successfully demonstrated the realization of multilevel conductance states in the FeFETs, which is important for amplifying effective bit density for embedded memory technologies. Notably, our FeFETs with In contact exhibits a larger read current of ~3.1 (5.2) μA in response to a 0.1 (1.0) ms programming pulse compared to those using Ti contacts, suggesting potential for low-power operation.

## RESULTS AND DISCUSSION

### FeFETs with different contact interfaces.

**Figure 1a** illustrates the cross-sectional schematic of the FeFET structure. A $MoS_2$ monolayer grown by metal-organic chemical vapor deposition (MOCVD) is utilized as the 2D channel and the AlScN layer, 45 nm thick, acts as an effective ferroelectric gate due to its high $P_r$ (~80-115 μC/cm$^2$).[16,17] The applied gate voltage ($V_g$) induces the switching of ferroelectric polarizations in AlScN, resulting in a change of the fermi level ($E_f$) of $MoS_2$. This modulation subsequently produces erased or programmed conductance states of the FeFET (**Figure 1b**). Here, metallic In has been chosen as contact electrodes to integrate with the $MoS_2$/AlScN structure with a low contact resistance. This is because In is known to be less chemically reactive with TMDs and has a work function ($W_F$ ≈4.1 eV) that is smaller than the electron affinity of monolayer $MoS_2$ ($\chi$ ≈4.4 eV), possibly resulting in the a negligible Schottky barrier height at the MSJ interface (i.e., $\Phi_B = W_F - \chi$ = -300 meV for ideal case)[20,22]. The device fabrication is scalable (see **Methods** for detail), which allows the successful fabrication of arrays of In-contacted $MoS_2$/AlScN FeFETs (**Figure 1c**). The device dimensions are a channel length ($L_{ch}$) of ~300 to 1,500 nm and a width ($W_{ch}$) of ~20 μm, as presented in a scanning electron microscopy (SEM) image (**Figure 1d**).

The typical transfer curve of an In-contacted FeFET, measured at $V_{ds}$ of 1 V in both semi-logarithmic and linear scales, is shown as red curves in **Figure 1e** and **1f**, respectively. The transfer characteristic of the FeFET displays a counterclockwise sweep direction with high gate-field modulation ($I_{on}/I_{off} > 10^7$) (**Figure 1e**), indicating the initiation of the ferroelectric switching mechanism. On the other hand, the $MoS_2$ FETs fabricated with a non-ferroelectric (non-Fe) dielectric gate ($SiO_2$) shows clockwise hysteresis caused by charge trap effects, further confirming the effect of AlScN (**Figure S1**). The memory window (MW), extracted by the threshold voltages ($V_{th}$) shift, is confirmed to be greater than ~7 V (**Figure 1f**). To prove the impact of In contacts, the FeFETs are also fabricated with Ti contacts (instead of In) for comparison, and the transfer curve results of the fabricated devices are displayed as blue curves in **Figure 1e** and **f**. Ti was selected for the comparison because it is chemically reactive when deposited on $MoS_2$, resulting in a degraded MSJ interface with high contact resistance.[20,25] As proof, a representative In-contacted $MoS_2$/AlScN FeFET shows a higher $I_{on}$ of ~100 μA/μm than that of Ti-contacted (~20 μA/μm) at $V_{ds}$ of 1 V. Such a high on-state current density in monolayer $MoS_2$ can also be realized in an FeFET with thicker AlScN (~100 nm), indicating the universality of the In contact for high resistance ratio switching (**Figure S2**). The representative output characteristics also display ~6 times higher saturation current in the FeFETs with In contacts over Ti contacts (**Figure 1g** and **Figure S3**). The saturated $I_{ds}$ of ~300 μA/μm is reached at only $V_{ds}$ of ~2.7 V for the FeFET with a 300 nm channel length, which is much higher than that of a Ti/$MoS_2$/AlScN FeFET (< 50 μA/μm). It is worth noting that the saturation $I_{ds}$ of our In-contacted FeFETs (~300 μA/μm) surpasses the highest $I_{ds}$ (~250 μA/μm) reported previously for Ti-contacted $MoS_2$/AlScN FeFETs with a 4-times shorter $L_{ch}$ of ~80 nm.[18]

To enhance the impact and reproducibility of our results, we investigate the performance of Ti and In contact electrodes on a large number of monolayer $MoS_2$/AlScN FeFETs, with over 20 devices tested for each type of contact rather than a comparison from a single representative I-V characteristic. The results are compiled and presented in the form of histograms (**Figure 1h-j**). The results show that devices with In contacts have ~3 times higher $I_{on}$ (~57.7 ± 18.3 μA/μm) and ~4 times better $I_{on}/I_{off}$ ($\sim 2 \cdot 10^7 \pm 1.5$) than devices with Ti contacts, when averaging 20 devices with $L_{ch}$ varied from ~300 to 1,500 nm (**Figure 1i, j**). On the other hand, the MW of the $MoS_2$/AlScN FeFET with In contacts (~4.5 ± 1.2 V) is smaller compared to the devices with Ti contacts (~5.4 ± 1.3 V) (**Figure 1h**). The smaller MW is a result of the negative shift of $V_{th}$ in the forward sweep, which is caused by the increased efficiency of current injection by the small-work-function In contact that makes it easier for the transistor to switch on.[26,27] The variability of $I_{on}$ with respect to $L_{ch}$ is characterized by calculating the coefficient of variation ($C_V = \sigma/\mu$, where $\sigma$ is the standard deviation and $\mu$ the mean value) for the In- and Ti-contacted $MoS_2$/AlScN FeFETs. The average $C_V$ for the In-contacted FeFETs is ~11.8 ± 6.2%, which is three times smaller than that for the Ti-contacted devices (~37.8 ± 20.7%) for the devices with the same $L_{ch}$ (~300 nm). The observed device-to-device variation is attributed to the non-uniformity of the contact and dielectric interface, as well as local charge inhomogeneities.[28]

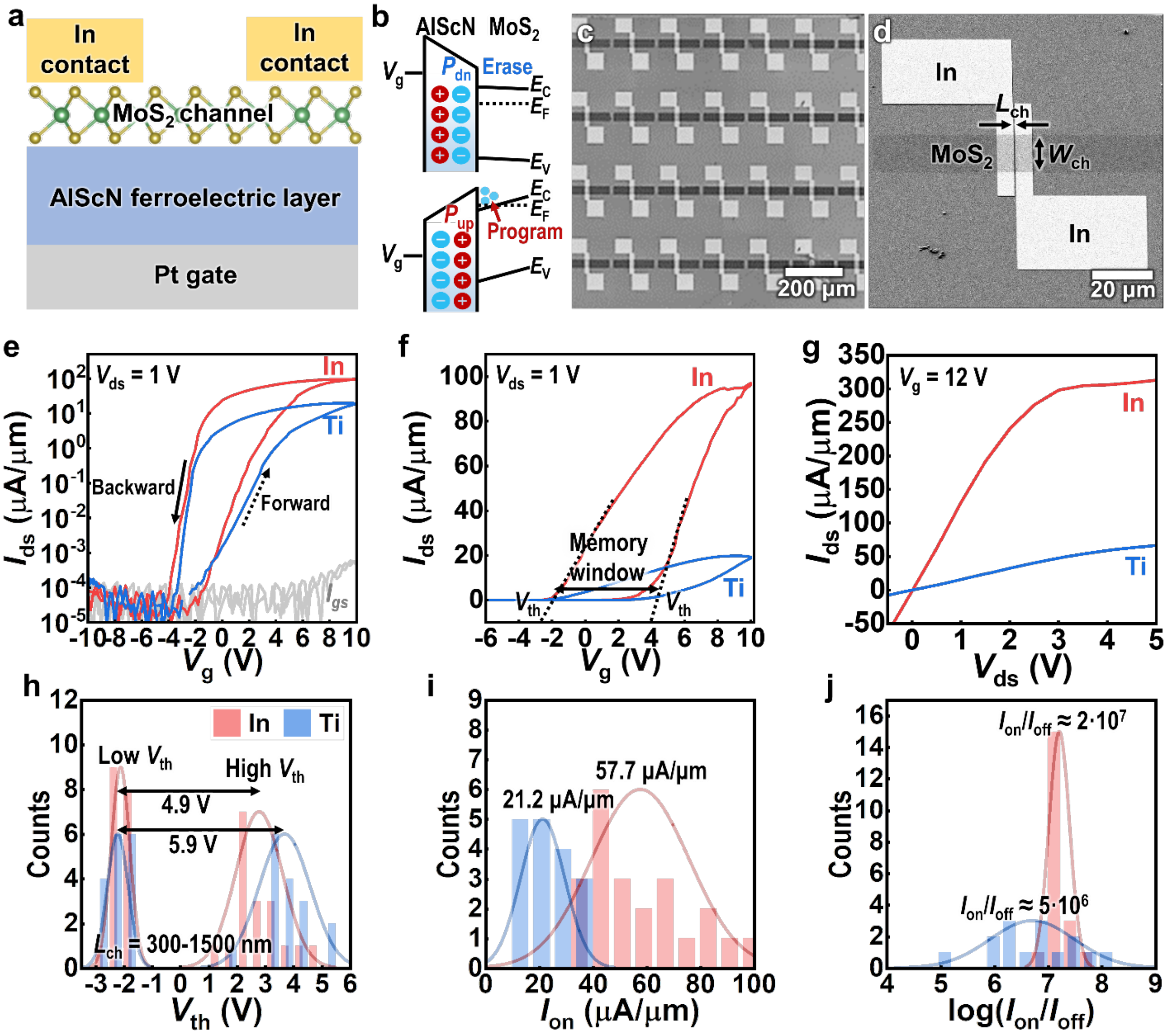

**Figure 1. Structure and electrical properties of a monolayer $MoS_2$/AlScN FeFET.** (**a**) Schematic diagram of a $MoS_2$/AlScN FeFET with In/Au contact electrodes. (**b**) Illustrations depicting the band bending of monolayer $MoS_2$, influenced by the polarization directions ($P_{up}$ or $P_{dn}$) of AlScN, which in turn govern the erase/program functions of the FeFET. The polarization bound charges with + or - symbols and mobile electrons are displayed in the bands. (**c, d**) SEM images of (**c**) the FeFET arrays and (**d**) the unit device with a channel width ($W_{ch}$) of 20 μm. (**e, f**) Transfer characteristics ($I_{ds}$-$V_g$) of the FeFET with In (red) and Ti contact electrodes (blue) under the $V_{ds}$ of 1 V in the (**e**) semi-logarithm and (**f**) linear scales. The transfer curves were recorded at a rate of 10 Hz with a $V_g$ spacing of 0.2 V. The differences in the subthreshold voltages ($V_{th}$) indicate the memory window (MW). (**g**) Output characteristic ($I_{ds}$-$V_{ds}$) of FeFET under the $V_g$ of 12 V (See Figure S3 for the dependence of output curves on $V_g$.). (**h-j**) Histograms of (**h**) $V_{th}$ values showing the memory windows, (**i**) on-state current density ($I_{on}$) at $V_g$ = 10 V, and (**j**) $I_{on}/I_{off}$ values depending on the In (red) and Ti contacts (blue). The performance metrics are extracted using the transfer curves under $V_{ds}$ of 1 V for devices with a channel length ($L_{ch}$) of 300-1,500 nm. $I_{on}/I_{off}$ is the maximum attainable ratio between $I_{on}$ and $I_{off}$.

**Transmission line method for FeFETs.**

To investigate the room-temperature electrical properties of the $MoS_2$/AlScN FeFETs, contact resistance ($R_c$) and channel's sheet resistance ($R_{ch}$) are extracted using the transmission line method (TLM). **Figure 2a** shows a plot of total resistance normalized by channel width ($RW_{ch}$) of the $MoS_2$/AlScN FeFETs with In and Ti contacts at $V_g$ of 10 V. The averaged $RW_{ch}$ increases linearly as a function of the $L_{ch}$ according to a relationship as follows:

$$RW_{Ch} = R_{ch} + 2R_c; \qquad (1)$$

$$R_{ch} = \frac{L_{ch}}{q \cdot n_{2D} \cdot \mu_{FE}} \qquad (2)$$

Here, $\mu_{FE}$ is field-effect mobility (=$(dI_{ds}/dV_g) \times (L_{ch}/W_{ch}C_iV_{ds})$, where $C_i = 2.5 \times 10^{-7}$ F/cm$^2$ is the dielectric constant of AlScN[16,17]). Using Equation 2, one can obtain $2R_c$ and $R_{ch}$ as they are the $y$-intercept and the slope of the $RW_{ch}$-$L_{ch}$ plot, respectively (**Figure 2a**). The calculated on-state $R_c$ value of In/$MoS_2$ MSJ FeFET approaches ~4.5 ± 1.9 kΩ·μm at $V_g$ of 10 V (average ± standard error from the linear fitting), which is almost two times smaller than that of Ti/$MoS_2$ MSJ FeFET (~8.8 ± 0.9 kΩ·μm) at the same $V_g$ (**Figure 2b**), representing better MSJ interface with In contact. Moreover, the TLM-extracted $R_{ch}$ of In-contacted $MoS_2$ FeFET decreases over $V_g$, reaching ~12.2 ± 2.2 kΩ/sq at 10 V (**Figure 2c**). The Ti/$MoS_2$ MSJ FeFET, meanwhile, demonstrates a higher on-state $R_{ch}$ (~19.9 ± 1.3 kΩ/sq). When comparing the smaller $R_c$ and $R_{ch}$ values with those of FeFETs employing Ti contact electrodes, the effectiveness of carrier transport through the In contact is evident (More explanation on how $R_c$ leads to a smaller $R_{ch}$ is provided in the following paragraphs for Figure 3.).

To evaluate the performance of our In-contact FeFETs, we have conducted a comprehensive comparison with other devices reported, considering parameters such as $R_{ch}$, $R_c$, $I_{on}$, $I_{on}/I_{off}$, and MW. Compared to the previous reports of $MoS_2$ FETs (with a non-Fe dielectric gate),[21-24,29,30] the on-state $R_{ch}$ of our FeFET is significantly smaller (> 11.4 kΩ/sq), even smaller than the quantum resistance (~$h/2e^2$) (**Figure 2d**). Even though it has slightly higher $R_c$ (~4.4-17 kΩ·μm) compared to the other studies on standard FETs[21-24,29,30] (~0.1-5 kΩ·μm), the $R_{ch}$ is lower, indicating significant conductance switching in our FeFETs. Furthermore, our FeFET outperforms all the previously reported $MoS_2$-based FeFETs,[10,11,31-35] as demonstrated by its improved $I_{on}$ (solid circles in **Figure 2e**), $I_{on}/I_{off}$, and MW values (**Figure 2f**). In addition, the achieved high $I_{on}$ in our device (~130 μA/μm) is higher than the previous report for $MoS_2$ FETs with In contact ($I_{on}$ ≈18 μA/μm)[22] and 2D metal-based contact electrodes ($I_{on}$ ≈0.1-3.4 μA/μm).[20,36,37] It is even higher than the FET with substitutionally doped channels ($I_{on}$ ≈0.05-4 μA/μm)[6,38] and comparable to FETs with semi-metal contacts of Sb ($I_{on}$ ≈124-569 μA/μm)[23] or Bi ($I_{on}$ ≈226 μA/μm)[21] that approach the quantum limit (see **Table S1** for more comparisons). Besides, our device has the highest $I_{on}/I_{off}$ ($\sim 2 \cdot 10^7$) and highest normalized MW (~0.11 V/nm; normalized by the ferroelectric thickness) among $MoS_2$-based FeFETs reported to date (**Figure 2f**),[10,11,31-35] which can provide a better read margin to minimize the error rate for memory applications.[39] The superior performance of our FeFET is not only due to the increased current injection efficiency by the In contact electrodes but also due to the high remnant polarization ($P_r$ ≈80-115 $\mu C/cm^2$) and large coercive field ($E_c$ ≈2-4.5 MV/cm) of the AlScN ferroelectric itself.[17] Theoretically, the $P_r$ and $E_c$ directly relate with the carrier density induced in the channel, $\Delta n_{2D,Fe}$ ($=P_r/q$), and the MW ($= 2E_c \times$ thickness of the ferroelectric), respectively.[19]

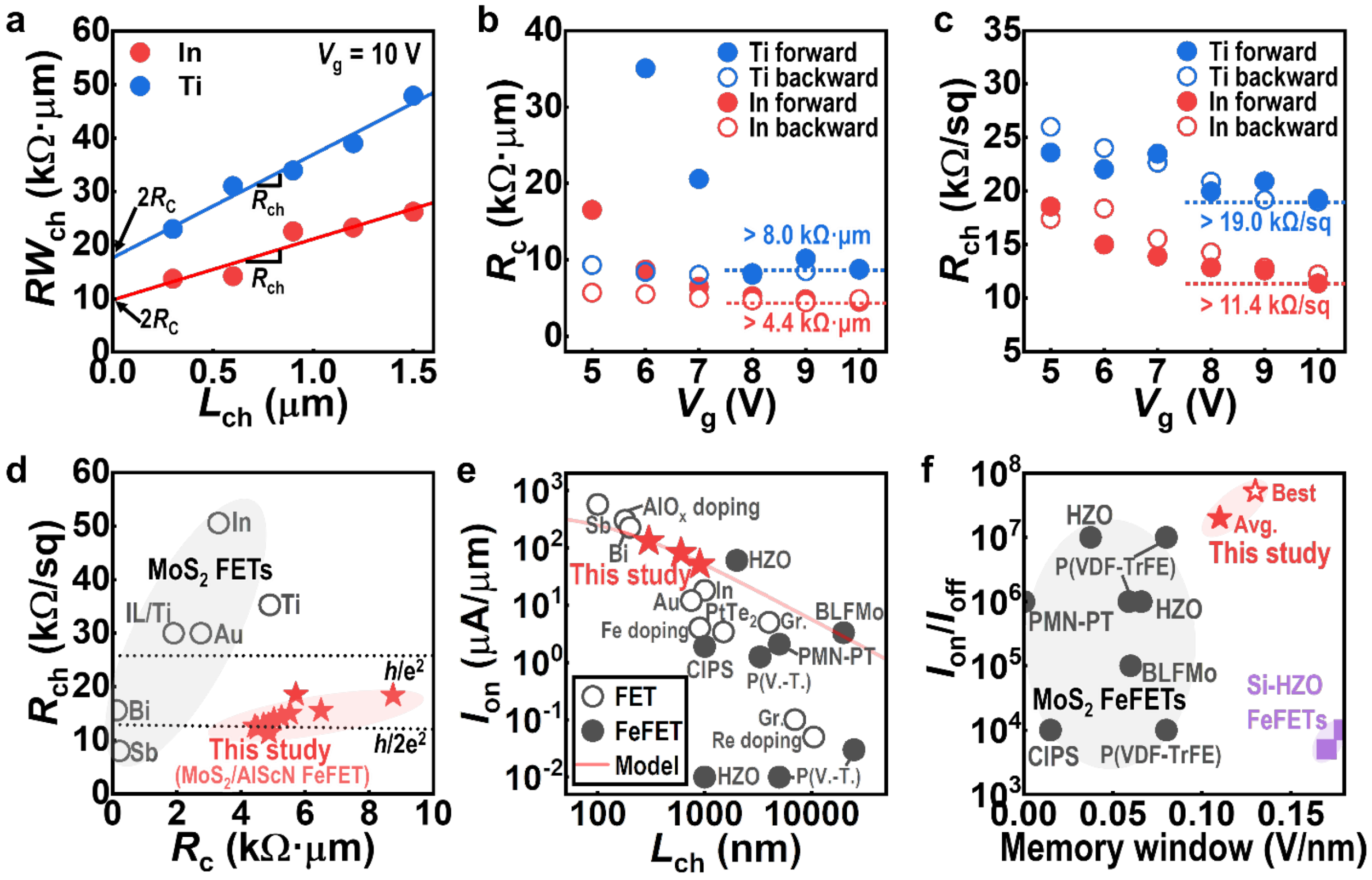

**Figure. 2. High resistance ratio switching in the $MoS_2$ FeFET with low contact resistance.** (**a**) Total resistance ($RW_{ch}$) of the FeFETs with In and Ti contact electrodes depending on the channel length ($L_{ch}$) measured at $V_g$ of 10 V during the forward sweep in transfer transport. The $y$-intercepts and the slope indicate the contact resistance ($2R_c$) and the channel sheet resistance ($R_{ch}$) of the TLM sets, respectively. (**b**) $R_c$ and (**c**) $R_{ch}$ of In- (red) and Ti-contacted $MoS_2$ FeFETs (blue) under the different $V_g$ in the forward (solid) and backward sweep (open). (**d**) Comparison of $R_{ch}$ as a function of $R_c$ of few-layer $MoS_2$ FETs[21-24,29,30] (circles) and our $MoS_2$/AlScN FeFET (stars). The contact electrodes for each FET are labeled. IL stands for interlayer contact. (**e, f**) Comparisons of (**e**) $I_{on}$ (normalized at $V_{ds}$ = 1 V) depending on the channel length ($L_{ch}$) and (**f**) $I_{on}/I_{off}$ and memory window (normalized by the thickness of ferroelectric) for our FeFET (red stars) with the literature values for $MoS_2$ FET[6,20-23,36-38,40] (open) and $MoS_2$ FeFET[10,11,31-35] (solid). Compared values are extracted from the DC measurement for the transfer characteristic of (Fe)FETs. The fitted curve in (e) reveals the ohmic relation of the $I_{on}$ values ($I_{on} \propto 1/L_{ch}$). For standard (non-ferroelectric) FETs, the contact electrodes are labeled; and for FeFETs, the ferroelectric material is labeled. The reported performances for the Si-channel HZO FeFETs[41,42] (purple squares) are also displayed in (f) to compare with 2D $MoS_2$-based FeFETs.

**Contact-limited mobility and carrier density.**

The current density of a transistor is directly related to mobility (or saturation velocity) and carrier density;[40] hence, to gain a better understanding of performance metrics in our devices, we examine them in this section. First, the two-terminal, field-effect linear mobility ($\mu_{FE}$) values of our device with $L$ = 1 μm are extracted (**Figure 3a**). The In-contact FeFET exhibits higher $\mu_{FE}$ than the Ti-contact device, attributed to its better contact interface and lower contact resistance. The degradation of a contact interface by Ti results in an increase in the contact resistivity ($\rho_C$) and the sheet resistance of $MoS_2$ underlying contact metals ($R_{sh}$; note that this is different from channel resistance, $R_{ch}$), as; $R_c \propto \sqrt{\rho_C R_{sh}}$,[43] as depicted in **Figure S4a, b**. On the other hand, soft metal In can minimize the degradation of $R_{sh}$ upon metal deposition because the structure of the underlying 2D $MoS_2$ can be preserved.[22,44] Similarly, the contact interfacial issue may reduce the intrinsic mobility of $MoS_2$ under the contact ($\mu_c$) by $R_{sh} = 1/qn_c\mu_c$, where the $n_c$ is the sheet carrier density of the semiconductor underneath the contact.[43]

We have also assessed the $V_g$-dependent carrier density ($n_{2D}$) of In-$MoS_2$/AlScN FeFETs in the on state by utilizing the TLM-extracted $R_{ch}$ and $\mu_{FE}$ values (**Figure 3b** and **Figure S5**). During the forward sweep, $n_{2D}$ in the 2D channel increases with $V_g$. At the moment the sweep direction changes from forward to backward, the $n_{2D}$ still increases due to the non-volatility of the ferroelectric polarization, reaching a maximum of ~$4.8\times10^{14}$ $cm^{-2}$, when the $V_g$ of 10 V gradually reduces to 9.4 V. As $V_g$ decreases further from 9.4 V, $n_{2D}$ decreases exponentially toward the off state (e.g., $n_{2D} \approx 2.6\times10^{12}$ $cm^{-2}$ at $V_g$ = 0 V). According to the standard parallel capacitance model (gray curves in **Figure 3b**), the $V_g$-induced $n_{2D}$ in the channel is much larger than what can be achieved solely with the AlScN dielectric gating (i.e., $\Delta n_{2D,cap} = C_i(V_g - V_{th})/q$ at $V_g > V_{th}$). This implies that the $P_r$ (= $\Delta n_{2D,Fe}/q$) of the ferroelectric AlScN plays a crucial role in the $V_g$-induced electron density in the 2D channel.

To determine the 2D carrier density induced solely by the ferroelectric polarization ($\Delta n_{2D,Fe}$), we subtracted the $\Delta n_{2D,cap}$ obtained using the capacitance model (gray curves in **Figure 3b**) from the calculated total $n_{2D}$ (data points in **Figure 3b**). We assume that the simplest case, wherein the increase in $I_{ds}$ is related to the capacitive and ferroelectric responses,[45-47] such that the total $n_{2D} = \Delta n_{2D,cap} + \Delta n_{2D,Fe}$. We then scale the resulting $\Delta n_{2D,Fe}$ values to calculate the ferroelectric polarization, $P_r$ values of the FeFET, as $P_r = q\cdot \Delta n_{2D,Fe}$ (shown in **Figure 3c**). For example, the maximum $\Delta n_{2D,Fe}$ value at $V_g$ of 9.4 V (~$4.6\times10^{14}$ $cm^{-2}$) corresponds to a $q\cdot\Delta n_{2D,Fe}$ value of ~73.7 $\mu C/cm^2$. Remarkably, the calculated $q\cdot \Delta n_{2D,Fe}$ value (~73.7 $\mu C/cm^2$ at 9.4 V) due to ferroelectric polarization in our device is much closer to the reported $P_r$ of our AlScN[16,17] ($P_r \approx$ 80-135 $\mu C/cm^2$). This indicates that ferroelectric gating in the In-contacted $MoS_2$/AlScN FeFETs is effectively supported by the high $P_r$ characteristic of AlScN. In addition, the result proves the $q\cdot\Delta n_{2D,Fe}$ values are also highly dependent on the contact properties. Using Ti as the contact electrode, instead of In, results in a maximum $q\cdot\Delta n_{2D,fe}$ value of only ~26 $\mu C/cm^2$ (**Figure 3d**). This suggests that the high $R_c$ of the MSJ interface hinders the utilization of the high $P_r$ inherent in AlScN[16,17] ($P_r \approx$ 80-135 $\mu C/cm^2$). When $R_c$ is high, the switching of the MFM capacitor under the contact may be imperfect, resulting in significantly lower partial polarization. The degraded $MoS_2$ layer (beneath the contacts) can create multiple trap sites, which may result in a depolarization field that prevents full $2P_r$ switching of AlScN.[35]

Given the huge mismatch of $P_r$ and carrier density in conventional 3D Si/$Hf_xZr_{1-x}O_2$ FeFETs, leading to substantial charge trapping at the Si/ferroelectric interface and reliability issues,[2] our findings suggest that a 2D channel FeFET with a superior contact interface can support the high carrier densities minimizing imbalance of the charges between channel and ferroelectric materials. The 2D $MoS_2$ channel with minimal oxidation,[18] compared to the Si with native oxide,[2,4,5,35] may enable the strong coupling between the ferroelectric polarization and the induced charge in the 2D channel due to the prevention of charge screening from the oxide layer. Moreover, the sub-1 nm thickness of 2D $MoS_2$ allows minimal effective body capacitance compared to a 3D semiconductor channel, which can lead to a smaller depolarization field.[48,49]

Therefore, combinations of high-$P_r$ AlScN with ohmic-contact 2D channel FeFETs can lead to significant conductance switching. To demonstrate this even more thoroughly, non-Fe FETs (i.e. standard FETs) with the exact same device configuration, fabrication process and material quality (including contact electrodes, channels, etc.), except for the gate dielectric (i.e., $SiO_2$ for FET), are fabricated for a comparison with AlScN/$MoS_2$ FeFETs. The representative transfer curves of the standard FET and FeFET are displayed in **Figure 3e**. Noticeably, the on-state current density of the FeFET is substantially superior to non-Fe FETs, more than an order of magnitude. Moreover, one can determine the $n_{2D}$ values in AlScN FeFETs by comparing the standard FETs' channel conductance ($G_{ch}$) and the "intrinsic" field-effect mobility ($\mu_i < \sim 1.4$ cm$^2$V$^{-1}$s$^{-1}$), extracted by the TLM method (**Figure 3f** and **Figure S5**). The on-state $n_{2D}$ in the AlScN FeFET calculated in this way is ~3.7 × 10$^{14}$ cm$^{-2}$, closely aligning with the value obtained in **Figure 3b** (~4.8×10$^{14}$ cm$^{-2}$), further proving its high doping density induced by the ferroelectric. Even in prior reports, such a high doping capacity in monolayer $MoS_2$ ($n_{2D} > \sim 10^{14}$ cm$^{-2}$) is challenging in traditional FETs with solid-state, non-ferroelectric gate dielectrics (**Figure S6a** and **Table S2**).

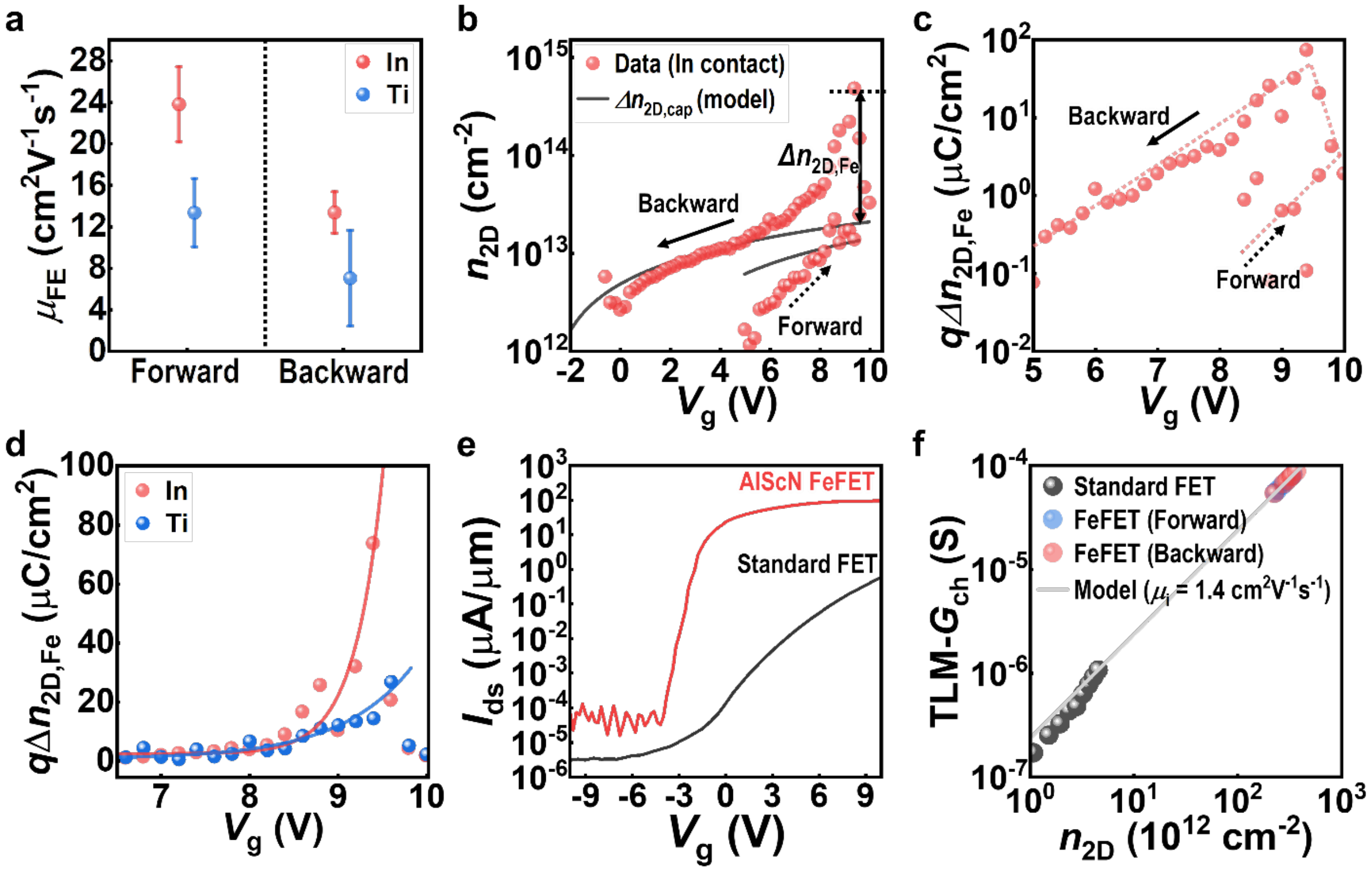

**Figure 3. Extraction of mobility and carrier density during the ferroelectric gating.** (**a**) Field-effect mobility ($\mu_{FE}$) of the $MoS_2$/AlScN FeFETs with In (red) and Ti contact electrodes (blue) depending on the sweep directions. (**b**) Extracted carrier density ($n_{2D}$) of the In-contact FeFET depending on $V_g$. The $\Delta n_{2D,cap}$ calculated by the parallel plate capacitance model is displayed as solid gray lines. (**c**) Electrical polarization ($P_r = q \cdot \Delta n_{2D,fe}$) calculated using the difference between $n_{2D}$ values ($\Delta n_{2D,fe}$) of the data and the capacitance model in (b). (**d**) Comparison of the representative $q \cdot \Delta n_{2D,fe}$ values of the In- and Ti-contacted FeFETs. The arrows indicate forward (dashed line) and backward (straight line) sweep directions during the transfer curve characterizations in Figure 1d, e. The lines in (c) and (d) are guides for the eyes. (**e-f**) Comparisons of the FeFET and standard (non-Fe) FET in terms of (**e**) transfer characteristic during the backward sweep, and (**f**) TLM-extracted channel sheet conductance ($G_{ch}$) as a function of $n_{2D}$. The standard FET has the same device configurations, dimensions, and materials as those in FeFET, except for the gate dielectric; i.e., $SiO_2$ instead of AlScN. The $n_{2D}$ of FeFET in (f) is calculated by assuming that the intrinsic field-effect mobility ($\mu_i$) calculated for the standard FET is comparable to that of FeFET.

**Low-temperature measurements.**

To further understand the transport mechanism, an $MoS_2$/AlScN FeFET with In contacts is additionally measured at various temperatures ($T$ = 13-300 K). **Figure 4a** demonstrates the transfer curves measured under the same $V_g$ sweep (-10 to 10 V) at $T$ of 240-300 K. One can observe that the turn-on voltage for electron transport increases, and the subthreshold swing (SS) gets steeper at the higher $T$. This is because the required $V_g$ to switch the polarizations in ferroelectric AlScN becomes higher at low $T$, resulting in the suppression of both the ferroelectric negative capacitance effect and hole conductance as $T$ decreases. Conversely, in the $T$ range of 13-120 K, the SS are almost comparable to each other, and the $V_{th}$ decreases at higher $T$ (**Figure 4b**). In this $T$ regime, the FET is no longer turned on/off by the ferroelectric switching but controlled by the purely capacitive response from the non-Fe dielectric, which is also indicated in the $I_{ds}$-$V_g$ hysteresis directions depending on $T$ in **Figure S7**. Consequently, in the subthreshold regime, the suppression of thermionic emission of the electrons over the Schottky barrier at the MSJ becomes more noticeable as the $T$ decreases, resulting in the $V_{th}$ and SS changes.

**Figure 4c** displays $T$-dependent channel conductance ($G$) at $V_g$ of -10 V (blue) and 10 V (red). First, we can observe that the slope of the curve changes around 200 K to some extent, indicating the transition from the non-Fe to Fe gating mechanism of the transistor. Besides, at $V_g$ of -10 V, the $G$ decreases when the $T$ decreases, as expected for temperature-dependent thermionic emission in insulators ($dR/dT > 0$, blue). Conversely, at a $V_g$ of 10 V, the $G$ improves when the $T$ declines ($dR/dT < 0$, red), probably due to the injected higher carriers, indicating its metallic phase. Even for monolayer $MoS_2$ transistors with strongly suppressed carrier scattering, a high carrier density greater than $n_{2D}$ ~$10^{13}$ $cm^{-2}$ (or $R_{ch} < 0.8h/e^2$; which is the case in our device as shown in **Figure 2d**) in a 2D channel is possible, which leads to a critical point that shows the transition between the insulating phase to the metallic phase.[50-52]

Accordingly, this metal-to-insulator (MIT) behavior does provide a good indication of gating-induced doping capacity to reach $n_{2D}$ ~$10^{13}$ $cm^{-2}$ (based on the theory[50-52]). We extract $dR/dT$ depending on $V_g$ at different $T$ regimes (**Figure 4d**), and here, the $V_g$ at which MIT occurs, or $V_{MIT}$, is the point at which $dR/dT$ changes from negative to positive. Noticeably, the $V_{MIT}$ is ~1 V at $T$ > 240 K, whereas $V_{MIT}$ is ~4 V for $T$ < 120 K, suggesting that Fe gating can support larger $n_{2D}$ than non-Fe gating and trigger band-like transport in the 2D channel.[50-53] Note that the $V_{MIT}$ is ~1 V at $T > 240$ K which is close to the voltage range where the $n_{2D}$ reaches the ~$10^{13}$ $cm^{-2}$ in **Figure 3b**, indicating their consistency.

The high $n_{2D}$-induced band-like transport is also supported by the observation of linear mobility evolution following the phonon-limited carrier transport model of $\mu \propto T^{-\gamma}$ with a $\gamma = 1.7$ for $T > 200$ K (**Figure 4e**). This extracted $\gamma$ value is comparable to the theoretical exponent $\gamma$ for monolayer $MoS_2$ ($\gamma = 1.7$) with the optical phonon scattering, not suppressed by other variations such as Schottky barriers.[50,53] The consistency of $T$-dependent $\mu_{FE}$ has been verified across three different devices in **Figure 4e**. In spite of device-to-device variations, all devices display carrier transport characterized by the same exponent $\gamma = 1.7$, maintaining consistency at $T = 200$-$300$ K, where ferroelectric gating affects the carrier injection. Given that this phonon-limited transport mechanism can be exclusively observed in $MoS_2$ FETs with dual gates or with ultra-low contact resistance,[23,50] our approach indicates that AlScN-based gating can provide a useful platform to enable high-mobility, high-carrier density 2D channel devices on a large scale. For example, $\mu_{FE}$ over 100 $cm^2V^{-1}s^{-1}$ is observed at low temperatures in two devices (i.e., ~114 and 116 $cm^2V^{-1}s^{-1}$ for Device 1 and 3, respectively, at 13 K in **Figure 4e**. Refer to comparisons with previous reports in **Figure S6b** and **Table S2**). It is worth noting that the field-effect mobility stays high at large carrier densities suggesting the ability of ferroelectric polarization to effectively screen charged impurity scattering from defects and interfaces. On the other hand, below 120-200 K, the mobility starts to saturate mostly limited by Coulomb scattering of defects. This saturation is also largely affected by the absence of a contribution from the ferroelectric switching to the carrier transport.

The effectiveness of transport through the AlScN-based FeFET with a high-quality contact interface is further confirmed by its smaller effective thermionic barrier height ($\Phi_B$) for electrons to overcome (**Figure 4f,g**), extracted by the following equation:

$$I_{ds} = \left[AA^{*}T^{3/2}\, exp\left(-\frac{q\Phi_B}{k_BT}\right)\right]\left[exp\left(\frac{qV_{ds}}{k_BT}\right) - 1\right] \qquad (3)$$

Here, $A$ is the junction area, and $A^*$ is the effective Richardson–Boltzmann constant. Hence, the value of $\Phi_B$ can be obtained from the slope of the linear fit to the Arrhenius plot ($ln$ ($I_{ds}/T^{3/2}$) *vs.*

$1/k_B T$) as depicted in **Figure 4f**. The calculated $\Phi_B$-$V_g$ plot suggests that $\Phi_B$ is substantially small even at -10 V (~79 meV) (**Figure 4g**). This is even smaller than those of the In (~375 meV)[22] and Bi contact (~130 meV),[21] and comparable to the Sb contact (~53 meV)[23] for $MoS_2$ FET in the completely off state. Moreover, at $V_g$ above -7 V (even in the subthreshold regime), the $\Phi_B$ values are negative (**Figure 4g** and its inset), indicating the large field emission, rather than thermionic emission, through the MSJ (as depicted in **Figure S4c, d**).

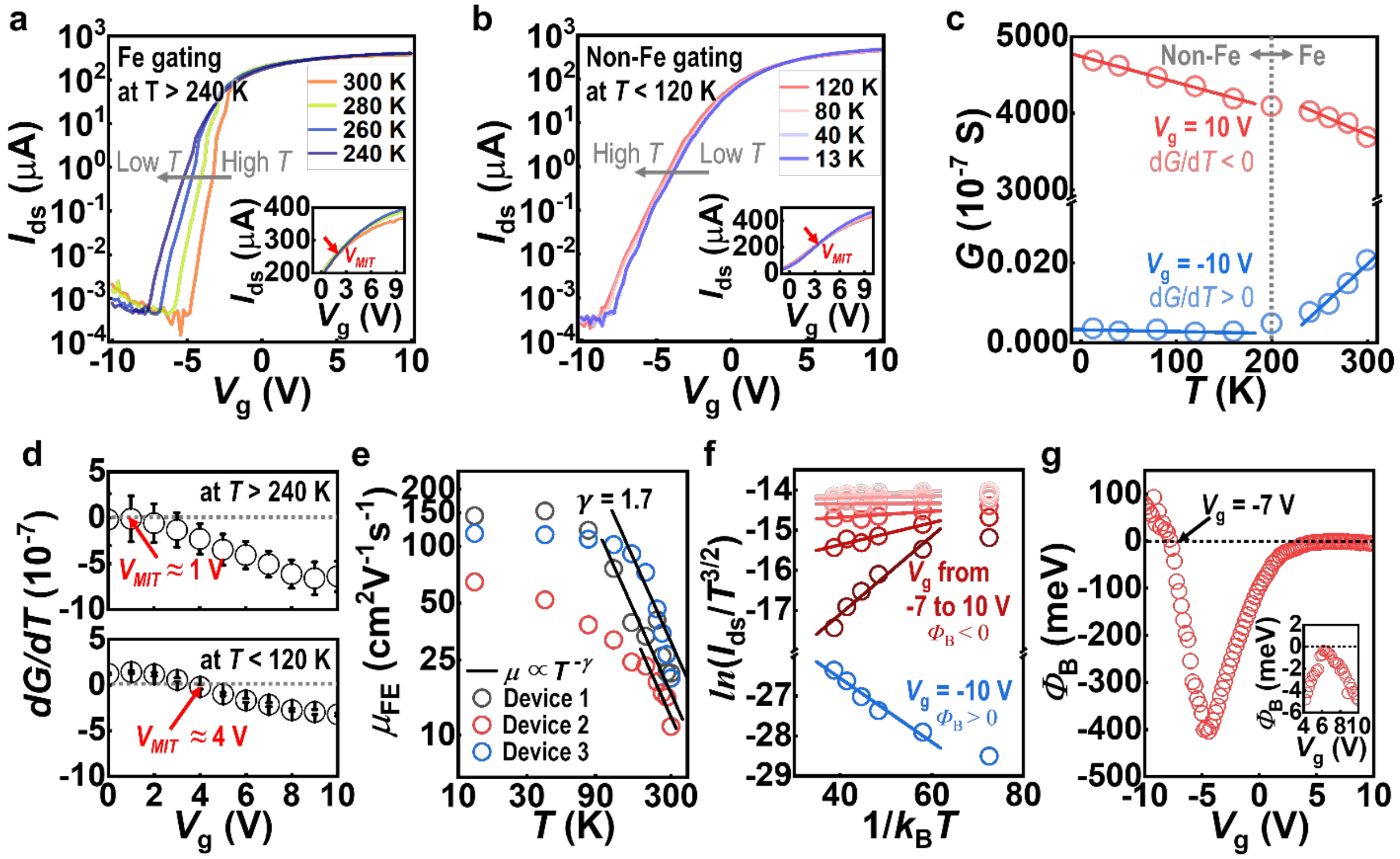

**Figure 4. Low-temperature measurements of FeFETs with its different gating mechanisms.** (**a, b**) Transfer curves of the same FeFET measured at temperatures of (**a**) $T$ = 240-300 K and (**b**) 13-120 K, during the backward sweep. Inset shows the zoomed-in plots in a linear scale near the $V_g$ range that shows metal-insulator electrical phase transition (MIT). (**c**) $T$-dependent channel conductance ($G$) at the applied $V_g$ of 10 (red) and -10 V. The negative slope ($dG/dT < 0$) indicates its metallic behavior whereas the positive slope ($dG/dT > 0$) implies its insulating property. (**d, e**) Extracted $dG/dT$ depending on the applied $V_g$, at (**d**) $T$ = 240-300 K and (**e**) 13-120 K. The specific $V_g$ that enables the MIT where the $dR/dT \approx 0$ is labeled as $V_{MIT}$. (**e**) $T$-dependent $\mu_{FE}$ of three FeFET devices, with the fitted phonon scattering model ($\mu \propto T^{-1.7}$). (**f, g**) Extraction of the effective thermionic barrier height ($\Phi_B$) at the metal-semiconductor junction between the $MoS_2$ and In contact. (**f**) Arrhenius plot ($ln(I_{ds}/T^{3/2}$ vs. $1/k_BT$) to calculate the $\Phi_B$ values The negative slopes indicate the absence of $\Phi_B$ (< 0), which is the case for $V_g$ measured at -7 to 10 V. (**g**) Extracted $\Phi_B$ depending on the $V_g$. The inset shows the zoomed-in curve for the $V_g$ range of -4 to 10 V, displaying the negative $\Phi_B$ for $V_g$ > 4 V.

**Pulse measurements.**

Lastly, we evaluate the resistive switching and the retention of In-contacted $MoS_2$/AlScN FeFET, by applying a voltage pulse at the gate ($V_g$). **Figure 5a** shows the time retention characteristics of binary states. Two distinct current states with a program/erase over $10^4$ at $V_g = 0$ V and $V_{ds} = 0.1$ V are achieved after poling the polarization of AlScN up and down by a 1 s program pulse $V_g$ of ± 12 V. It is found that varying the program pulse amplitude induces various $I_{ds}$, offering the possibility of multilevel storage in the FeFET, beyond simple binary states. **Figure 5b** (and **Figure S8**) shows the multilevel conductance stored after each $V_g$ pulse with different amplitude from -12 to 12 V (with 0.5 V differences) and a width of 1 ms. The different $I_{ds}$ are recorded depending on the $V_g$, and the programming of $I_{ds}$ is possible under a $V_g$ bigger than 6.5 V. The stepwise program state is clearly distinct for a total of 58 levels stored at a $V_g > 6.5$ V with increasing pulse numbers, promising its 5-bit operation.

The programming speeds of FeFET are further estimated by measuring the $V_g$ pulse-width dependence of the $I_{ds}$. **Figures 5c, d** illustrate the detailed program characteristics of the FET examined by varying pulse amplitude ($V_g$ ranging from 4 to 12 V) and width ($t_{width}$ of 0.1 and 1.0 ms). As the $V_g$ amplitude increases, the corresponding $I_{ds}$ increase gradually as a response from the AlScN ferroelectric polarization switching, where $I_{ds}$ is higher for In-$MoS_2$ FeFETs than Ti-contacted ones in both $t_{width}$ cases contributed from its larger $n_{2D}$ induced at the same $V_g$. This suggests that In-contact FeFETs can use a lower pulse amplitude or smaller pulse width compared to Ti-FeFETs to complete the switching for the same program state, demonstrating the great potential of FeFETs for low-power and high-speed non-volatile memory applications.

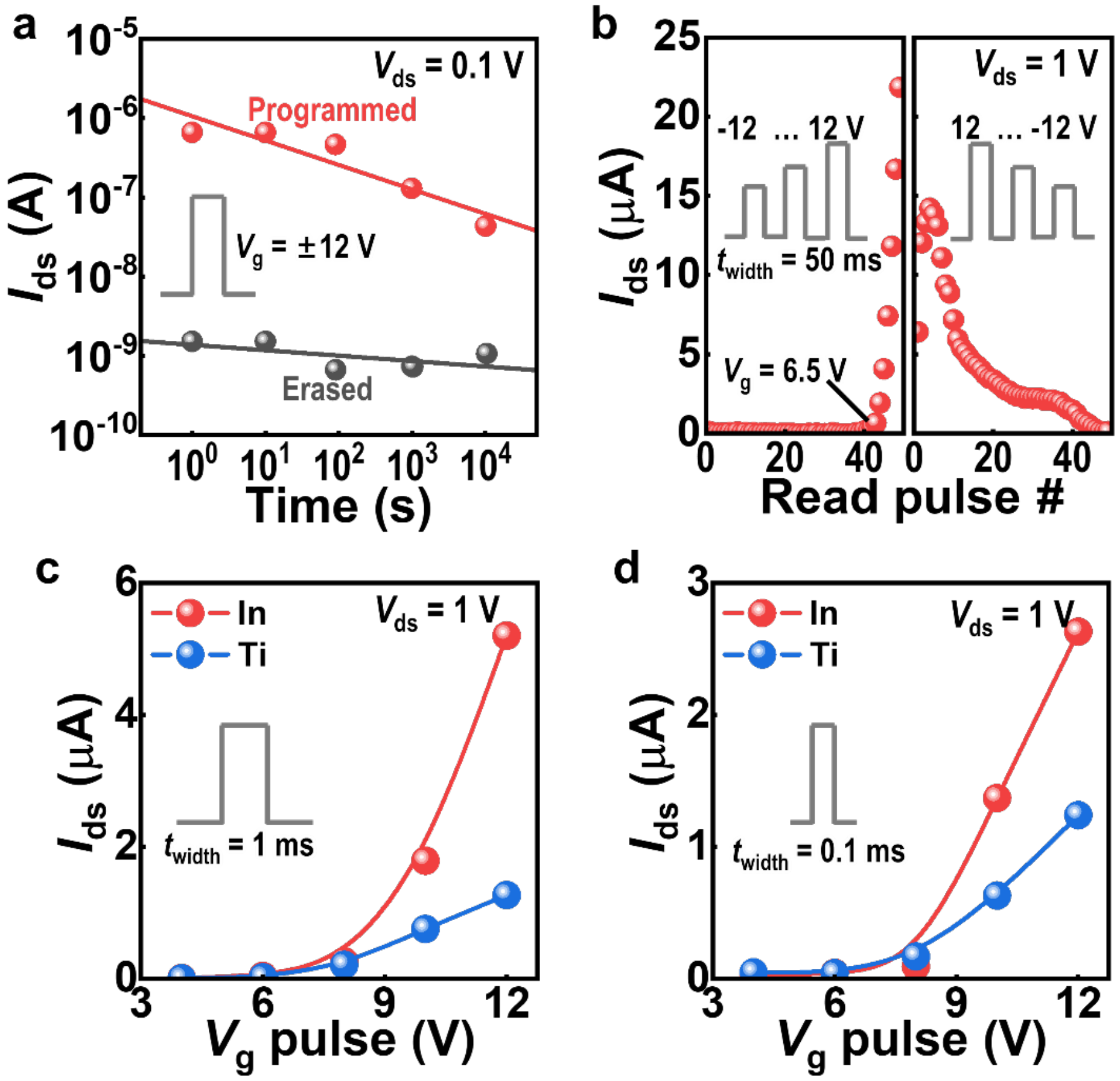

**Figure 5. Memory retention and programming speed characterizations at room temperature.** (**a**) Retention properties of the $MoS_2$/AlScN FET with In contact electrodes, at the program and erase states in response to $V_g$ pulses with an amplitude of 12 V and width of 1 s. $I_{ds}$ values were recorded at $V_{ds}$ = 0.1 V and $V_g$ = 0 V. (**b**) Multiple $I_{ds}$ states programmed/erased under the pulse train with different amplitudes from $V_g$ = -12 to 12 V with a pulse width ($t_{width}$) of 50 ms. $I_{ds}$ were probed at $V_{ds}$ = 1 V and $V_g$ = 0 V after applying for program/erase pulse. (**c**, **d**) Program state currents of the FeFETs with different contact electrodes, as a function of $V_g$ pulse amplitude. The programming $t_{width}$ was 1 ms and 0.1 ms in the case of (c) and (d), respectively.

## CONCLUSION

In summary, the impacts of contact resistance on the monolayer $MoS_2$/AlScN FeFETs are investigated with In contact electrodes. High saturated current density (~320 μA/μm at $V_{ds}$ = 3 V), $I_{on}/I_{off} > 10^7$, and normalized MW > 0.11 V/nm are achieved in the FeFET. This achievement is attributed to the low contact resistance at the In/$MoS_2$ MSJ interface down to ~4.4 kΩ·μm, coupled with high remnant polarization, $P_r$, of AlScN. Consequently, a significantly large carrier density approaching ~$4.8\times10^{14}$ $cm^{-2}$ is obtained, corresponding to a partially switched $P_r$ value of ~73.7 μC/$cm^2$. Furthermore, the high doping concentration induced by the ferroelectric allows for observation of nearly ideal phonon-limited transport (with an exponent $\gamma$ of 1.7) with field-effect mobilities > 100 $cm^2V^{-1}s^{-1}$ at cryogenic temperatures. In addition, pulse measurements conducted in FeFETs demonstrate the realization of 58 conductance levels. We have observed that the stored $I_{ds}$ values in In-contacted FeFETs surpass those of Ti-contacted devices under the same program pulse amplitude and width. Therefore, our findings present a reproducible approach for achieving the high-current-density and high-mobility 2D channel AlScN FeFETs. Moreover, the low contact resistance and reconfigurable operations of the studied FeFETs make them suitable for integration into low-power electronics, offering potential additional functionalities.

## METHODS

### Device fabrication.

$Al_{0.68}Sc_{0.32}N$ with a thickness of 45 nm was deposited on pre-deposited Pt using a pulsed-DC reactive sputtering deposition system (Evatec, CLUSTERLINE 200 II pulsed DC PVD).[16,17] The co-sputtering was conducted using separate 4-inch Al and Sc targets, at a chuck temperature of 350 °C, with 10 sccm of Ar gas flow and 25 sccm of $N_2$ gas flow under the constant pressure of ~$1.45 \times 10^{-3}$ bar, which is the same as that utilized in our previous reports for FeFETs.[15,18] The preferentially oriented (111) Pt layer promoted the growth of AlScN with highly textured ferroelectricity along the [0001] direction. MOCVD-grown monolayer $MoS_2$ thin film was then transferred from its growth substrate of sapphire to as-grown AlScN using KOH-based wet transfer. A polymeric supporting layer of polymethyl methacrylate (PMMA) was coated on the $MoS_2$/sapphire and then floated onto 0.1 M KOH solution to detach $MoS_2$ from the sapphire substrate. The detached PMMA/$MoS_2$ free-standing layer was rinsed in distilled ionized water several times to avoid any damage to AlScN from the KOH-based residues. After the transfer of PMMA/$MoS_2$ to AlScN, the PMMA was removed by acetone for 10 min. The integration process for the monolayer $MoS_2$/AlScN FeFETs necessitates the prevention of any high-energy process, thermal budget above 180 °C, and exposure to basic solution to ensure high-quality FeFETs. Source and drain contact of In/Au (10/40 nm) were defined through e-beam lithography (Elionix ELS-7500EX) followed by thermal evaporation (Kurt J. Lesker Nano-36). In particular, the deposition rate for In metal should be adjusted to below 0.03 Å/s to minimize any high-energy damage to the $MoS_2$ surface during the thermal evaporation and to pursue small contact resistance[22]. It should be noted that, in this work, the Ti- and In-contacted FeFET arrays were fabricated on the same AlScN substrate with the same MOCVD-grown $MoS_2$ film, thereby there were fewer variations in the materials and the channel/dielectric interface. The channel-defining process was conducted using the reactive ion etcher (RIE) with oxygen gas (Jupiter II RIE Plasma Etcher). The Pt beneath the AlScN was used as a gate contact, where the gate contact region was exposed via the wet etching of AlScN on the substrate by dipping it in the KOH solution after the fabrication of the In/$MoS_2$/AlScN/Pt heterostructure. No additional heat treatment was conducted after the device fabrication.

**Device characterization.**

Room-temperature electrical measurements were conducted in the air at ambient temperature in a Lakeshore probe station using a Keithley 4200A semiconductor analyzer. Low-temperature electrical measurements were carried out by a Keithley 4200A semiconductor parameter analyzer in a closed-cycle cryogenic probe station (Advanced Research Systems, Inc.) with a base pressure of $\sim 10^{-6}$ torr. All statistically estimated values in the paper are reported as "average ± standard deviation," unless otherwise specified. SEM was used to capture images at 5 keV with the backscattered detector (TESCAN S8000X).

## ASSOCIATED CONTENT

Supporting information with Figure S1-8 and Table S1-2

## AUTHOR INFORMATION

### Author Contributions.

S. Song and K.-H. Kim contributed equally to this work.

S. S., K.-H. K. and D. J. conceived the experiments and wrote the manuscript with input from all authors. S. S., R. K., and M. D. conducted the low-temperature measurements. N. T., C. C., and J. M. R. grew the $MoS_2$. R.K. and M.D. assisted with cryogenic transport measurements. J. Z., and R. H. O. conducted the AlScN deposition and characterization.

## CONFLICT OF INTEREST

The authors declare no conflict of interest.

## DATA AVAILABLITY STATEMENT

The data that support the findings of this study are available from the corresponding author upon reasonable request.

## ACKNOWLEDGMENT

D.J. and S.S. acknowledge primary support from the Office of Naval Research (ONR) Nanoscale Computing and Devices program (N00014-24-1-2131). D.J. and K.H-K acknowledge partial support from the Air Force Office of Scientific Research (AFOSR) GHz-THz program grant number FA9550-23-1-0391. S.S. acknowledges partial support for this work by Basic Science Research Program through the National Research Foundation of Korea (NRF) funded by the Ministry of Education (Grant No. 2021R1A6A3A14038492). M.D. and R.K. acknowledge funding from DOE grant DE-SC0023224. A portion of the sample fabrication, assembly and characterization were carried out at the Singh Center for Nanotechnology at the University of Pennsylvania, which is supported by the National Science Foundation (NSF) National Nanotechnology Coordinated Infrastructure Program grant NNCI-1542153. The MOCVD $MoS_2$ samples were grown in the 2D Crystal Consortium Materials Innovation Platform (2DCC-MIP) facility at Penn State which is supported by the National Science Foundation under cooperative agreement DMR-2039351.

Supporting information

# High-Performance Ferroelectric Field-Effect Transistors with Ultra-High Current and Carrier Densities

Seunguk Song,[1,6] Kwan-Ho Kim,[1,6] Rachael Keneipp,[2] Nicholas Trainor,[3] Chen Chen,[4] Jeffrey Zheng,[5] Joan M. Redwing,[3,4] Marija Drndić,[2] Roy H. Olsson III,[1] and Deep Jariwala[1,*]

[1]*Department of Electrical and Systems Engineering, University of Pennsylvania, Philadelphia, Pennsylvania 19104, United States*

[2]*Department of Physics and Astronomy, University of Pennsylvania, Philadelphia, Pennsylvania 19104, United States*

[3]*Department of Materials and Science and Engineering, Pennsylvania State University, University Park, Pennsylvania 16801, United States*

[4]*2D Crystal Consortium Materials Innovation Platform, Materials Research Institute, Pennsylvania State University, University Park, Pennsylvania, 16801 United States*

[5]*Department of Materials Science and Engineering, University of Pennsylvania, Philadelphia, Pennsylvania 19104, United States.*

[6]*These authors equally contributed to this work.*

[*]Author to whom correspondence should be addressed: dmj@seas.upenn.edu

- Figures S1-8
- Table S1-2
- Supporting References

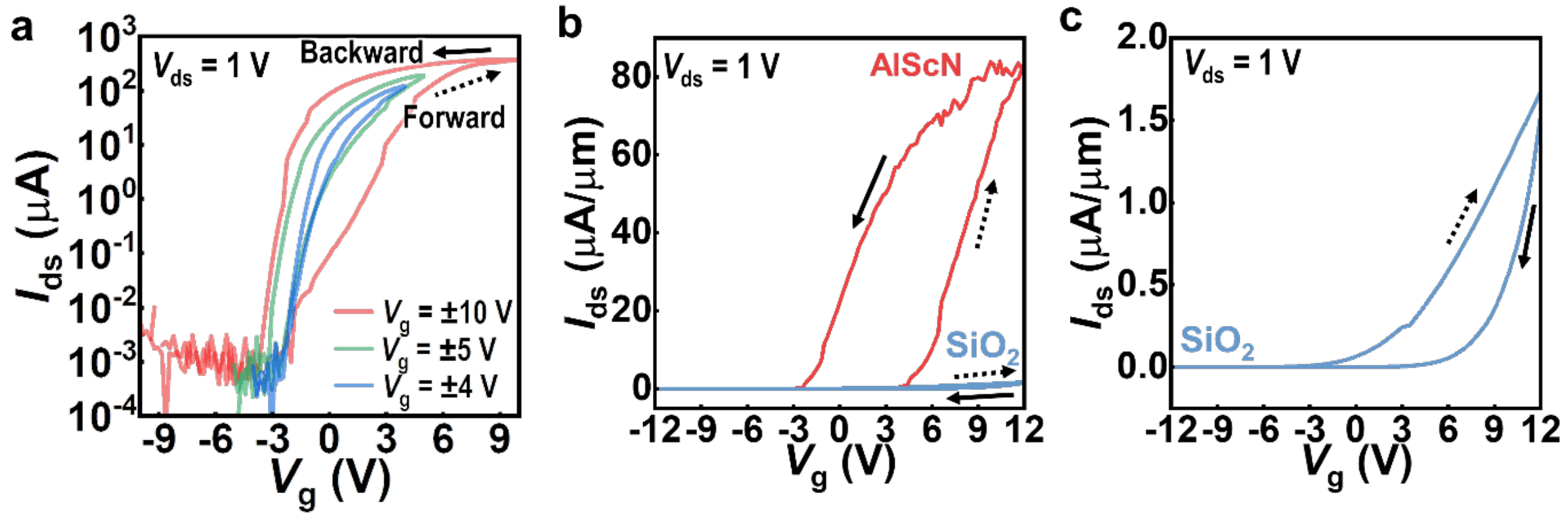

**Figure S1. Ferroelectric switching mechanism of $MoS_2$/AlScN FeFET with In contacts.** (**a**) Transfer characteristics for various $V_g$ sweep ranges (±4, ±5, and ±10 V), all showing counterclockwise hysteresis, as indicated by the arrows. (**b**) Comparison of the $I_{ds}$-$V_g$ plots for $MoS_2$ FeFET (red) and standard FET (blue) with $V_g$ applied through AlScN (45 nm thickness) and $SiO_2$ (50 nm thickness) gate substrates. Both devices were subjected to a $V_g$ sweep up to ±12 V for fair comparison. (**c**) Zoomed-in transfer curve of the $MoS_2$ FET gated with $SiO_2$ non-FE dielectric, from panel (b). The arrows indicate the hysteresis directions.

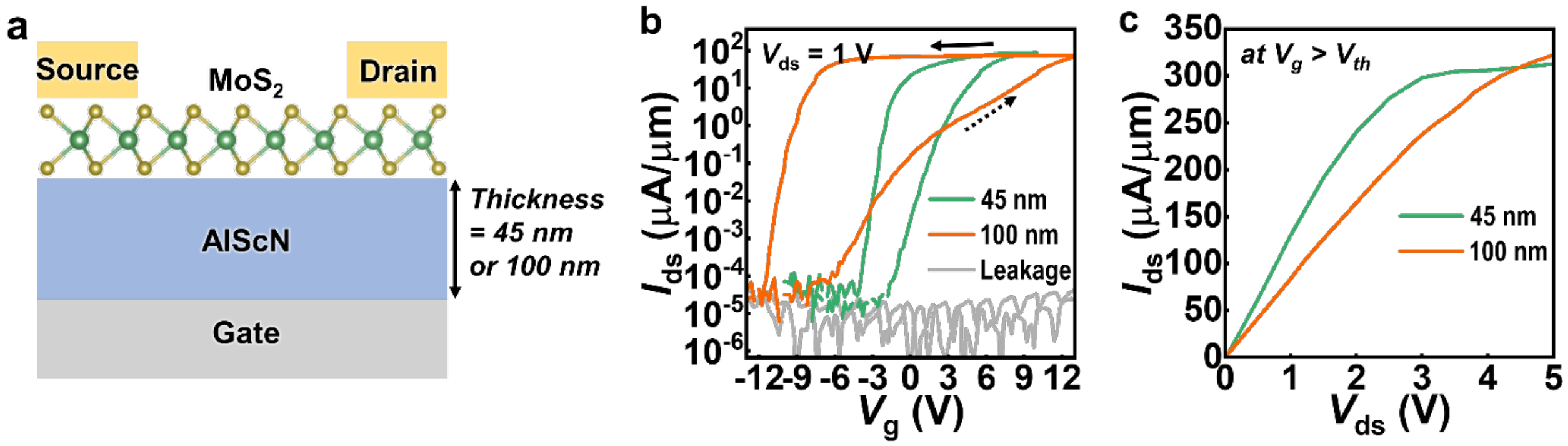

**Figure S2. High-performance of FeFET with a thicker AlScN ferroelectric.** (**a**) Cross-sectional schematic of the device structure. (**b**) Transfer and (**c**) output curves of the 100 nm- and 45-nm thick AlScN-based FeFETs. The output curves are measured under the maximum $V_g$ in (b), which is 10 and 14 V for 45 and 100 nm AlScN case, respectively.

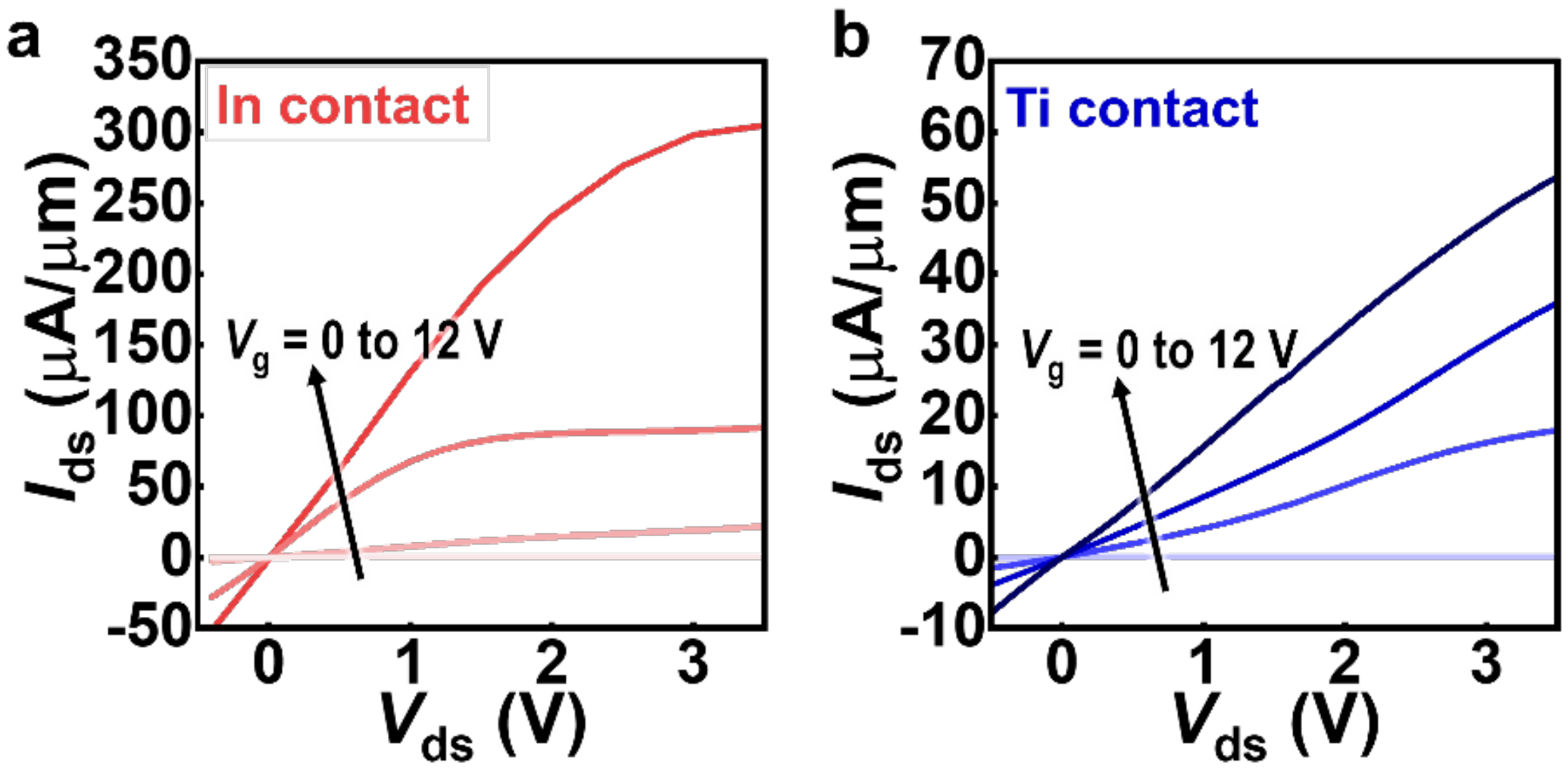

**Figure S3**. **Gate-dependent output characteristics of FeFETs.** The output curves ($I_{ds}$-$V_{ds}$) of $MoS_2$/AlScN FeFET with (**a**) In and (**b**) Ti contact electrodes. In contact FeFET shows more than five times higher $I_{ds}$ than Ti contact FeFET when $V_{gs}$ is 12 V.

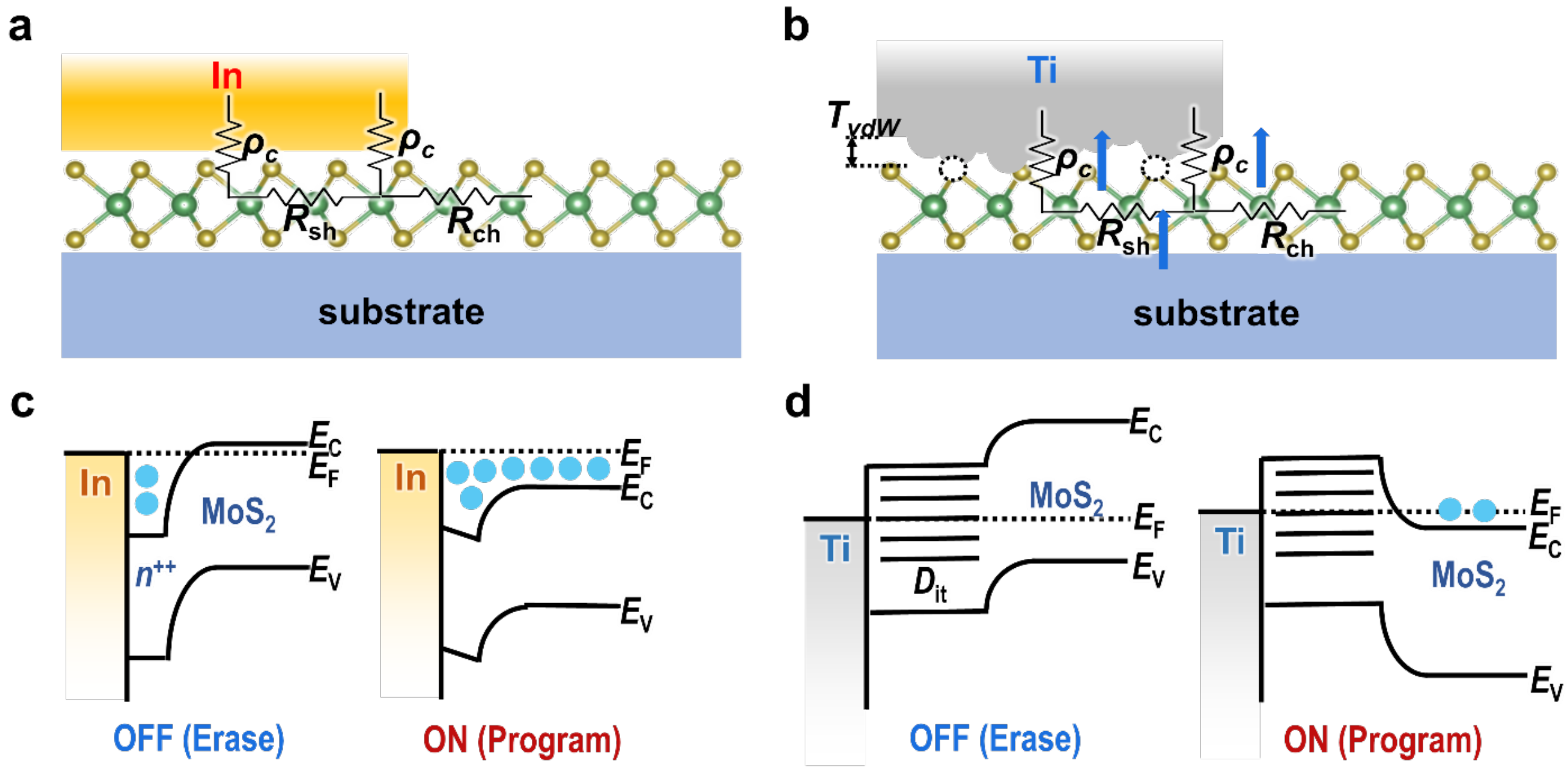

**Figure S4. Contact interface effects on device performance.** (**a, b**) Illustration depicting the relationship between contact resistivity ($\rho_C$) and sheet resistance of $MoS_2$ underlying contact metals ($R_{sh}$), highlighting the degradation of the contact interface by Ti. (**c, d**) Schematic representations of MSJ barrier heights and width depending on the contact interface. The schematic in (c) illustrates the significantly small thermionic barrier height ($\Phi_B$) at the In-$MoS_2$ MSJ even at -10 V (~79 meV; erase state), possibly caused by the degenerately doping by the carrier transfer from In to $MoS_2$ ($n^{++}$). The diagram on the right in (c) depicts the $E_F$ above $E_c$, leading to the dominance of field emission over thermionic emission through the MSJ, which is completely different from the Ti-$MoS_2$ MSJ case in (d).

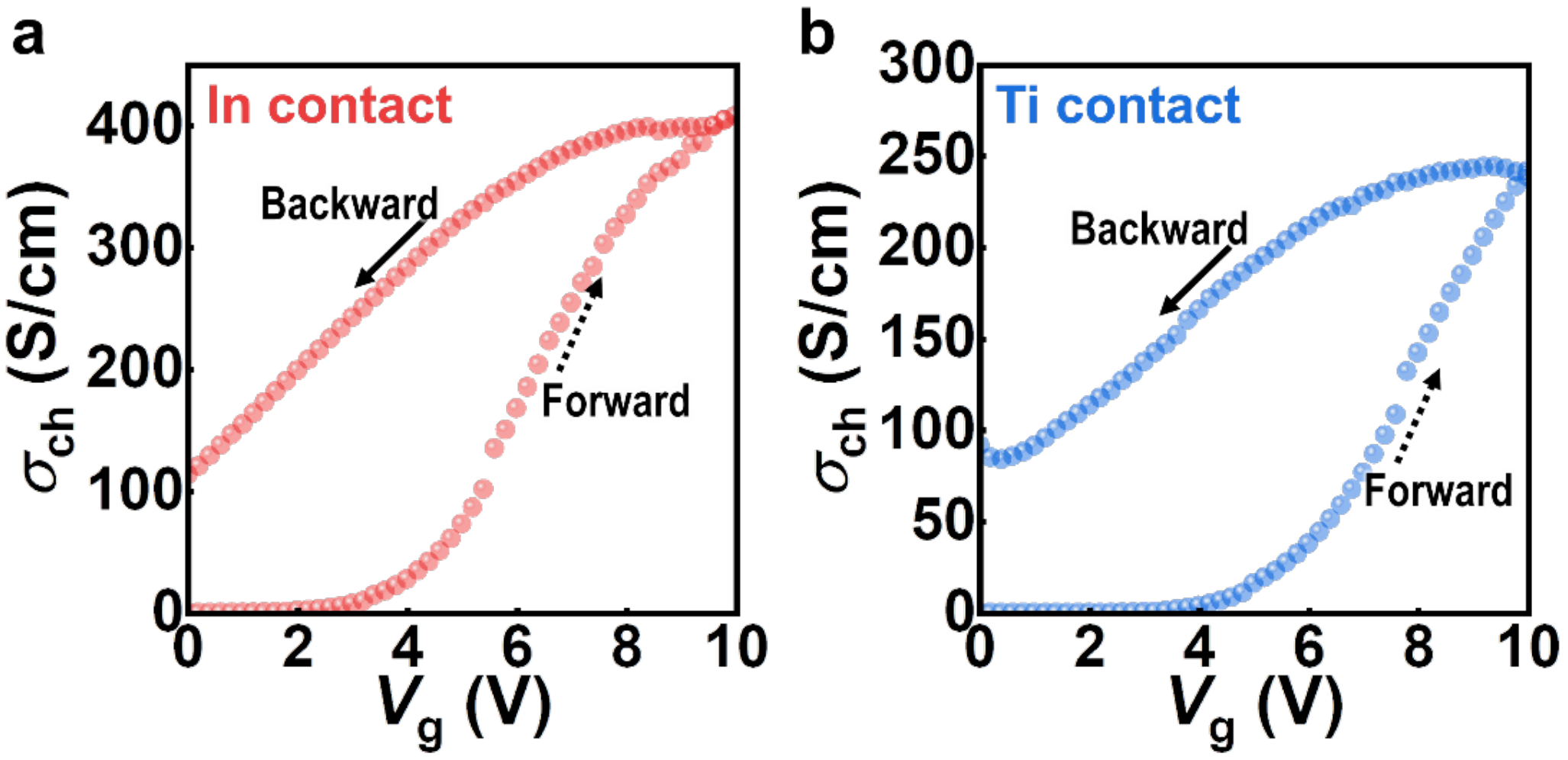

**Figure S5. TLM-extracted channel resistivity ($\sigma_{ch}$) of the $MoS_2$/AlScN FeFETs measured at $V_{ds}$ = 1 V.** The $\sigma_{ch}$ varied depending on (**a**) In and (**b**) Ti contact**.** The $\sigma_{ch}$ are extracted by excluding $V_g$-dependent $R_c$ in their total $R$, according to Equation (1) and Equation (2) in the main text.

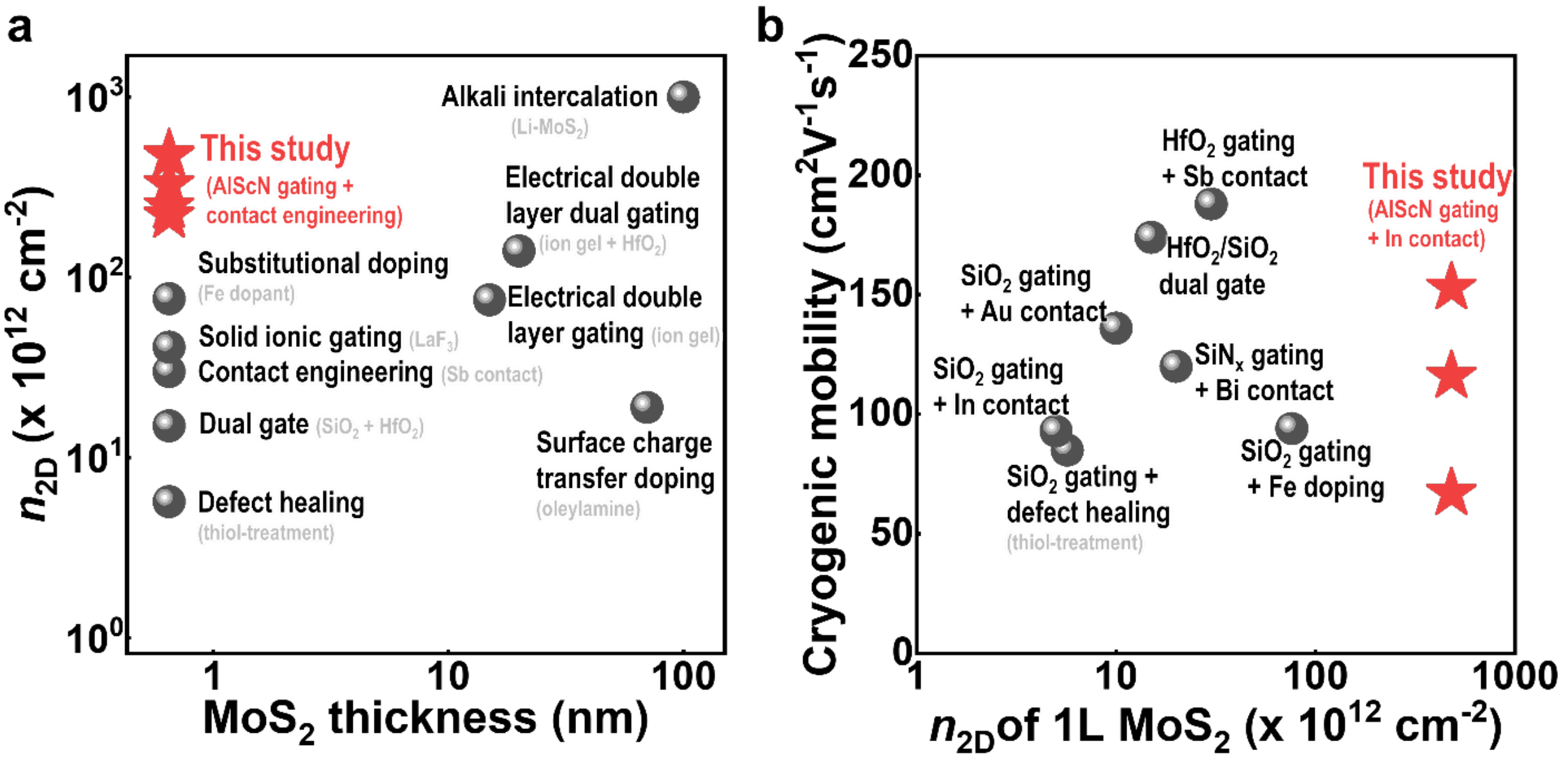

**Figure S6. Benchmark plots of different doping methods and gate field effects on $MoS_2$.** (**a**) Comparison of carrier densities ($n_{2D}$) in $MoS_2$ with different thicknesses, under different doping methods, or with gating methods. This highlights the challenge of achieving high doping levels ($n_{2D}$ > ~$10^{14}$ $cm^{-2}$) in a thinner $MoS_2$ as reported in previous studies.[1-10] (**b**) Charge density ($n_{2D}$) *vs*. cryogenic mobility for different gate dielectric layers in transistors with a monolayer (1L) $MoS_2$ channel.[5,6,8,9,11-13] Each point on the plot is labeled according to the specific gate dielectric type used in the measurements. The dielectric constants ($k$) for $SiO_2$, $HfO_2$, and AlScN are 3.9, 25, and 14, respectively.[14,15] Note that the dual-gated[9] and defect-healed[8] $MoS_2$ are mechanically exfoliated, while others are grown using CVD.[5,6,11-13]

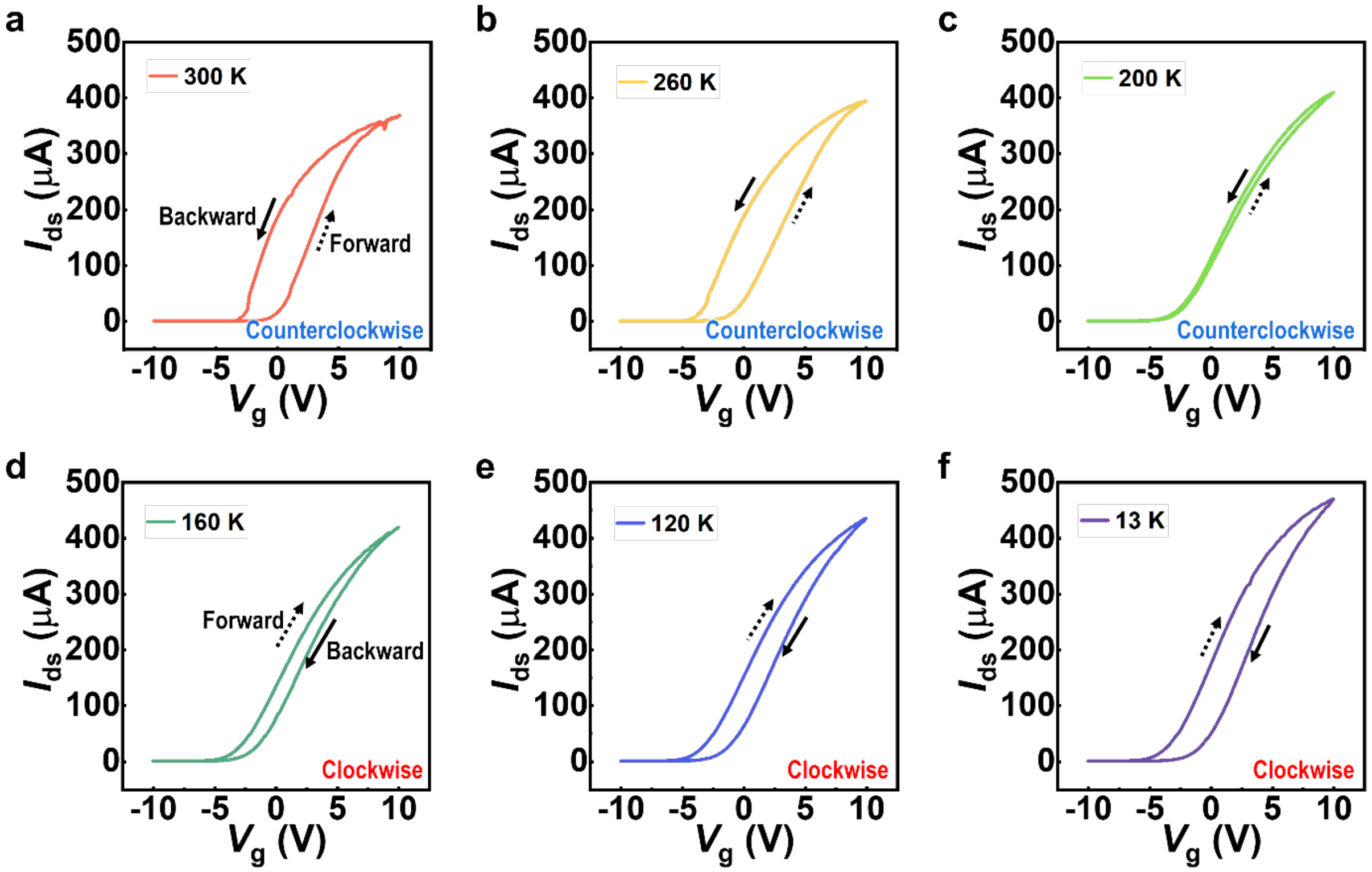

**Figure 7. Transition of FET behavior with various *T*.** The conductance switching by field effect transitions from Fe gating to gating by non-Fe dielectric capacitance, with corresponding changes in $I_{ds}$-$V_g$ hysteresis directions, depending on *T*, i.e., counterclockwise direction at (**a**) 300 K, (**b**) 260 K, and (**c**) 200 K, and clockwise direction at (**d**) 160 K, (**e**) 120 K, (**f**) 13 K.

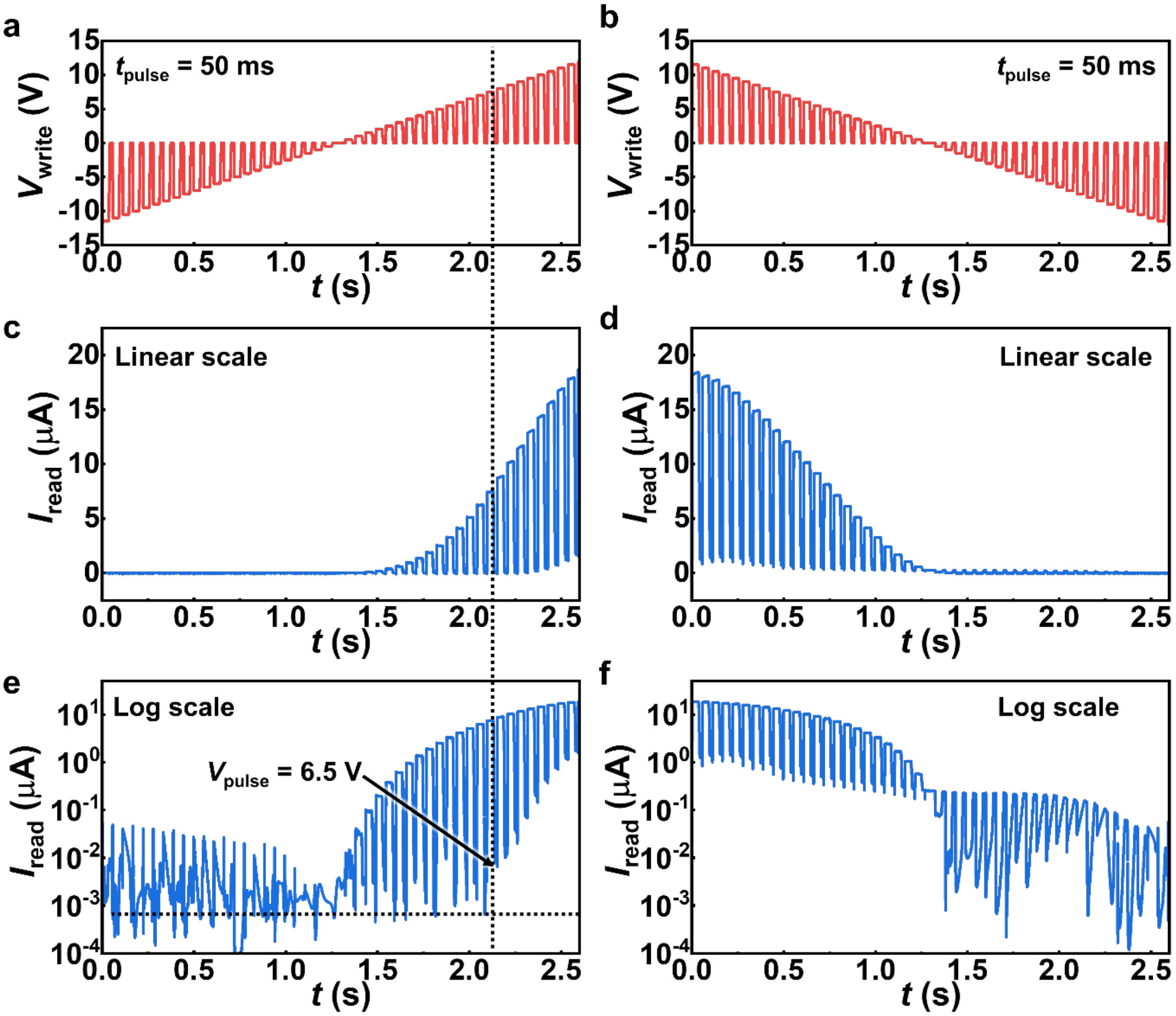

**Figure S8. Evaluation of $I_{ds}$ reconfigurability in In-contacted $MoS_2$/AlScN FeFET.** Multilevel conductance stored after each $V_g$ pulse, as depicted in (**a, b**), with varying amplitudes from -12 to 12 V (with 0.5 V differences) and a width of 50 ms. The read $I_{ds}$ are depicted in (**c, d**) linear and (**e, f**) log scale. The recording of different $I_{ds}$ depending on $V_g$ indicates the feasibility of programming $I_{ds}$ under a $V_g$ greater than ~6.5 V. A stepwise program state is clearly discernible for a total of 58 levels stored at $V_g$ > 6.5 V with increasing pulse numbers, suggesting its 5-bit operation potential.

**Table S1. Comparisons of *n*-type (FE)-FETs with 2D semiconductor channels.** For FeFET with 2D channel, the 3D ferroelectric gate of 0.7Pb($Mg_{1/3}Nb_{2/3}$)$O_3$-0.3Pb$TiO_3$ (PMN-PT), $Bi_{0.85}La_{0.15}Fe_{0.92}Mn_{0.08}O_3$ (BLFMO), $CuInP_2S_6$ (CIPS), poly(vinylidene fluoride-co-trifluoroethylene) (P(VDF-TrFE)), $Hf_{1-x}Zr_xO_2$ (HZO) were used.

| Device | Channel | Gate dielectric or ferroelectric | Contact metal or doping method | $R_c$ (kΩ·μm) | $n_{2D}$ ($10^{12}$ $cm^{-2}$) | $L_{ch}$ (nm) | $\mu_{FE}$ ($cm^2V^{-1}s^{-1}$) | | $I_{on}$ (μA·μm) at $V_{ds}$ = 1 V | $R_{ch}$ (kΩ/sq) | MW (V/nm) | $I_{on}/I_{off}$ ratio | Ref. |
|---|---|---|---|---|---|---|---|---|---|---|---|---|---|
| | | | | | | | at RT | at cryogenic *T* | | | | | |
| **FeFET** | **CVD-$MoS_2$** | **AlScN** | **In/Au** | **4.43** | **480** | **300** | **23.8** | **153** (at 30 K) | **130 (313 at $V_{ds}$ = 5 V)** | **12.2** | **0.11** | **$10^7$** | **This study** |
| FET | CVD-$MoS_2$ | $SiO_2$ | Au + $AlO_x$ doping | 0.48 | 20 | 180 | 33.5 (intrinsic) | N/A | 300 | 8.1 | N/A | $10^6$ | 16 |
| | CVD-$MoS_2$ | $SiO_2$ | $PtTe_2$ | 168 | 9 | 1,500 | 9.7 | N/A | 3.4 | 36.1 | | $10^7$ | 17 |
| | CVD-$MoS_2$ | $SiN_x$ | Bi | 0.123 | 20 | 200 | 55 | 120 (at 77 K) | 339 at $V_{ds}$ = 1.5 V | 17.9 | | $10^7$ | 11 |
| | CVD-$MoS_2$ | $SiO_2$ | In | 3.3 | 5 | 1,000 | 28 (at 280 K) | 93 (at 77 K) | 18 | 50.6 | | $10^6$ | 12 |
| | CVD-$MoS_2$ | $HfO_2$ | Sb | 0.209 | 30 | 1,000<br>400<br>100 | 87 (at 320 K) | 188 (at 50 K) | 124<br>268<br>569 | 8 | | $10^7$ | 5 |
| | CVD-$MoS_2$ | $HfO_2$ | Graphene | 59 | 16.57 | 7,000 | 17 | N/A | 0.1 | N/A | | $10^7$ | 18 |
| | CVD-$MoS_2$ | $SiO_2$ | Graphene | 115 | 3 | 4,000 | 35 | N/A | 5 | N/A | | $10^8$ | 19 |
| | CVD-$MoS_2$ | $SiO_2$ | Re doping | 26.25 | 6.5 | 10,500 | N/A | | 0.05 | N/A | | N/A | 20 |
| | CVD-$MoS_2$ (monolayer) | $SiO_2$ | UHV Au | 2.75 | 10 | 750 | 33 (intrinsic) | 136 (at 81 K, intrinsic) | 12 | 30 | | N/A | 13 |
| | CVD-$MoS_2$ | $SiO_2$ | Fe doping | 0.67 | 76 | 900 | 54 | 94 (at 100 K) | 4 | N/A | | $10^8$ | 6 |
| | Exfoliated $MoS_2$ (monolayer) | $SiO_2$ | Sn | 200 | 2 | N/A | 60 | 105 (at 1.4 K) | N/A | 60 | | N/A | 21 |
| | CVD-$MoS_2$ | $SiO_2$ | CuS | 27.3 | N/A | 2000 | N/A | | 1 | 545 | | $10^6$ | 22 |
| | CVD-$MoS_2$ | $SiO_2$ | Ti/Au + Low-T growth | 4.9 | 6.1 | 100 | 33 | N/A | 140 | 35.3 | | $10^7$ | 23 |
| | Exfoliated few-layer $MoS_2$ | hBN/$SiO_2$ | $MoSe_2$/Ti/Au | 1.9 | 7.9 | 400 | 50 | 160 (at 160 K) | 59 | 30 | | $10^6$ | 24 |
| | Exfoliated few-layer $MoS_2$ | $LaF_3$ gating (solid superionic conductor) | | N/A | 41 | N/A | 60 | 532 (at 2 K) | N/A | 2 | | $10^4$-$10^5$ | 10 |
| FeFET | Exfoliated InSe | CIPS | In/Au | N/A | | 10,000 | 728 | N/A | 0.14 | N/A | 0.03 | $10^6$ | 25 |
| | CVD-$MoS_2$ | $C_{60}$-doped P(VDF-TrFE) | Cr/Au | | | 25,000 | 30.2 | N/A | 0.006 at $V_{ds}$ = 0.2 V | | 0.059 | $10^6$ | 26 |
| | Exfoliated $MoS_2$ (13 nm) | $Al_2O_3$/BLFMo | Au | | | 20,000 | N/A | | 3.3 | | 0.060 | $10^5$ | 27 |
| | Exfoliated $MoS_2$ (6 nm) | $Al_2O_3$/HZO | Ti/Au | | | 2,000 | N/A | | 30 at $V_{ds}$ = 0.5 V | | 0.0375 | $10^7$ | 28 |
| | CVD-$MoS_2$ | PMN-PT | Ni/Au | | | 5,000 | N/A | | 2.08 | | 0.0002 | $10^6$ | 29 |
| | Exfoliated $MoS_2$ (7 nm) | CIPS | Ni | | | 1,000 | N/A | | 0.19 at $V_{ds}$ = 0.1 V | | 0.015 | $10^4$ | 30 |
| | Exfoliated $MoSe_2$ | P(VDF-TrFE) | Ti/Au | | | 5,000 | 6.77 | N/A | 0.001 at $V_{ds}$ = 0.1 V | | 0.08 | $10^4$ | 31 |
| | Exfoliated $MoS_2$ (4.62 nm) | P(VDF-TrFE) | Cr/Au | | | 3,300 | 95.6 | N/A | 2.5 at $V_{ds}$ = 2 V | | 0.08 | $10^7$ | 32 |
| | CVD-$MoS_2$ | HZO | Ti/Au | | | 1,000 | N/A | | 0.01 at $V_{ds}$ = 1 V | | 0.066 | $10^6$ | 33 |

**Table S2. Comparisons of doping strategies for 2D $MoS_2$.**

| Doping strategy | Channel | Gate | Contact metal | Maximum $n_{2D}$ ($10^{12}$ $cm^{-2}$) | $\mu_{FE}$ ($cm^2V^{-1}s^{-1}$) | | $I_{on}$ (μA·μm) at $V_{ds}$ = 1 V | $I_{on}/I_{off}$ ratio | $R_{ch}$ at RT (kΩ/sq) | Scattering coefficient (γ) | Ref. |
|---|---|---|---|---|---|---|---|---|---|---|---|
| | | | | | at RT | at cryogenic *T* | | | | | |
| **AlScN gating** | **1L-$MoS_2$ (CVD)** | **AlScN** | **In/Au** | **480** | **23.8** | **153** (at 30 K) | **130 (313 at $V_{ds}$ = 5 V)** | **$10^7$** | **11.3 (TLM) 0.36 ($L_{ch}$ = 300 nm)** | **1.69** | **This study** |
| Electrical double layer gating | 15 nm $MoS_2$ | Ion gel | Au | 75 | N/A | 86 (at 220 K, Hall device) | 0.85 | $10^2$ | 1.2 | N/A | 1 |
| | 20 nm $MoS_2$ | Ion gel (DEME-TSFI) and $HfO_2$ dual gate | N/A | 1.4 | N/A | 240 (at 20 K, Hall device) | N/A | N/A | 2 (at 100 K) | N/A | 2 |
| Alkali intercalation | ***Bulk*** $MoS_2$ with K, Cs, Li | N/A | N/A | 400-1000 | N/A | | N/A | N/A | N/A | N/A | 3,4 |
| Contact engineering | 1L-$MoS_2$ (CVD) | $HfO_2$ | Sb | 30 | 87 (at 320 K) | 188 (at 50 K) | 569 | $10^7$ | 8 | 1.8 | 5 |
| Substitutional doping | Fe-doped $MoS_2$ | $SiO_2$ | Cr/Au | 76 | 54 | 94 (at 100 K) | 4 | $10^8$ | N/A | N/A | 6 |
| Surface charge transfer doping | Oleylamine-doped 70 nm $MoS_2$ | $SiO_2$ | Au | 19 | 25 | N/A | 11.8 | 2.9 | N/A | N/A | 7 |
| Defect healing | Thiol-treated 1L-$MoS_2$ | $SiO_2$ | Ti/Pd | 5.7 | ~30 (81 with four terminal device) | ~85 at 10 K (& 320 with four terminal) | N/A | N/A | 16.9 | 0.72 | 8 |
| Dual gate | 1L $MoS_2$ | $SiO_2$ and $HfO_2$ | Au | 15 | N/A | 174 (at 4 K with four terminal device) | N/A | N/A | 4.3 | 1.4 | 9 |
| Solid superionic conductor gate | Few-layer $MoS_2$ | $LaF_3$ gating | Au | 41 | 60 | 532 (at 2 K) | N/A | $10^4$-$10^5$ | 2 | N/A | 10 |

## Supporting References